# Superwind evolution: the young starburst-driven wind galaxy NGC 2782

Jimena Bravo-Guerrero[1], Ian R. Stevens[2]
[1] Universidad Iberoamericana, Prol. Paseo de la Reforma 880, Lomas de Santa Fe, D.F., 01219, Mexico
[2] School of Physics and Astronomy, University of Birmingham, Edgbaston, Birmingham, B15 2TT, UK
(E-mail: jimena.bravo@gmail.com, irs@star.sr.bham.ac.uk

**ABSTRACT**

We present results from a 30 ksec *Chandra* observation of the important starburst galaxy NGC 2782, covering the 0.3–10 keV energy band. We find evidence of a superwind of small extent, that is likely in an early stage of development.

We find a total of 27 X-ray point sources within a region of radius $2D_{25}$ of the galaxy centre and which are likely associated with the galaxy. Of these, 13 are ULXs ($L_X \geqslant 10^{39}$ erg s$^{-1}$) and a number have likely counterparts. The X-ray luminosities of the ULX candidates are $1.2 - 3.9 \times 10^{39}$ erg s$^{-1}$. NGC 2782 seems to have an unusually large number of ULXs.

Central diffuse X-ray emission extending to $\sim 3$ kpc from the nuclear region has been detected. We also find an X-ray structure to the south of the nucleus, coincident with H$\alpha$ filaments and with a 5 GHz radio source. We interpret this as a blow-out region of a forming superwind. This X-ray bubble has a total luminosity (0.3–10 keV) of $5 \times 10^{39}$ erg s$^{-1}$ (around 15% of the total luminosity of the extended emission), and an inferred wind mass of $1.5 \times 10^6$ M$_\odot$.

We also discuss the nature of the central X-ray source in NGC 2782, and conclude that it is likely a low-luminosity AGN (LLAGN), with a total X-ray luminosity of $L_X = 6 \times 10^{40}$ erg s$^{-1}$, with strong Fe line emission at 6.4 keV.

## 1 INTRODUCTION

Starbursts are brief ($\sim 10^8$ yr) episodes of intense star formation in galaxies. In starburst galaxies, the star formation rate is high enough to consume the star-forming gas on a relatively short timescale. Starbursts produce important effects on their environments, and can influence the structure, evolution and formation of galaxies, by returning energy and heavy element-enriched gas to their surroundings via outflows from the starburst stellar populations (Weedman et al. 1981; Moorwood 1996).

Starburst-driven galactic winds evolution can be briefly described as follows: when the kinetic energy of the collective effect of the supernovae and stellar wind has been efficiently thermalised via shocks, a high pressure zone of hot gas will expand, sweep up interstellar material and form an X-ray emitting bubble, with low density and temperatures of $T \sim 10^7 - 10^8$ K (Chevalier & Clegg 1985). This bubble will carry on expanding, sweeping up more ambient medium, forming a superbubble, which will continue to expand until its swept-up shell fragments due to Rayleigh-Taylor instabilities, blowing out the hot gas into the intergalactic medium. This way, the superbubble in most star-forming galaxies results in a bipolar superwind, perpendicular to the disc, that blows out some of the ISM of the galaxy and much of the energy and material ejected by massive stars in the starburst (Heckman et al. 1993; Strickland & Stevens 2000). Therefore, galactic winds are considered the primary mechanism of energy and metal enrichment of the intergalactic medium (Veilleux et al. 2005).

Theoretical and hydrodynamic models (Chevalier & Clegg 1985; McCarthy et al. 1987; Heckman et al. 1993; Mac Low & Ferrara 1999; Strickland & Stevens 2000; Sofue & Vogler 2001; Strickland et al. 2004a,b; Veilleux et al. 2005; Cooper et al. 2009; Creasey et al. 2013) have been developed over the years for galactic superwinds; they agree with the conceptual model described above, and been validated observationally by several multi-wavelength studies of superwind galaxies. The observed morphology of superwinds can be quite different in different galaxies and will depend on the specific conditions within the galaxy. We see structures ranging from nuclear superbubbles to a biconical structures with filamentary morphology, and the scales can range from $\sim 1$ to 20 kpc.

The models mentioned above have focussed on the thermal expansion of hot gas generated by the supernova explosions in the starburst, to drive the galactic scale outflows. Recently, Thompson et al. (2015) have considered the role that radiation pressure on dust can have on outflows from galaxies (both starburst dominated and AGN dominated). If dusty material, in a cloud or a shell, is optically thick to UV radiation then the material can be driven to large velocities by continuum absorption of radiation (with the resulting velocity exceeding that of the escape speed of the galaxy). Thompson et al. (2015) discuss whether such a model can be applicable to the M82 outflow and concluded that it was probably not responsible for the dusty material seen at very large distances above the galaxy disc (though the situation was not completely clear and depended on the past luminosity of M82).

Spectacular examples of fully fledged superwinds include M82 (Strickland et al. 1997; Stevens et al. 2003; Strickland & Heckman 2009), NGC 253 (Strickland & Stevens 2000), NGC 3256 (Lehmer et al. 2015), and NGC 3079 (Heckman et al. 1990; Veilleux et al. 1994; Cecil et al. 2001; Strickland et al. 2003) but the earlier development of superwinds, particularly as the superbubble reaches the blow-out phase, has received less attention. The subject of this study, NGC 2782, may well be an example of a galaxy in such a phase and we aim to describe the detailed mechanisms of emission and physical state of the gas in this galaxy.

NGC 2782 (Arp 215), classified as peculiar $SABa(rs)$ (Sandage & Tammann 1981), also $Sa(s)$ Peculiar (de Vaucouleurs et al. 1991), is a nearby galaxy ($D = 37$ Mpc) that encompasses several important astrophysical issues. It also presents several characteristics that suggest a starburst-driven outflow in an early stage of evolution.

NGC 2782 harbours a central and circumnuclear starburst, powered by a gas-rich nuclear stellar bar, with a star-formation that has been estimated to be around 3–6 $M_\odot$ yr$^{-1}$ (Jogee et al. 1999). The galaxy's disturbed morphology suggests that it is the product of a collision between two galaxies or a merger remnant and the starburst may have been triggered by the interaction (Smith 1994). NGC 2782 shows a pair of H I tails, the eastern tail is optically brighter than the western tail, and three 'ripples' at radii of 25, 40 and 60 arcsec have been seen. There are also several H II regions in the eastern tail (Smith et al. 1999) as well as a bright H II region in the western tail, as recently discussed by Knierman et al. (2012). Saikia et al. (1994) presented a 5 GHz radio continuum map that displays well-defined structures to the north and south of the nucleus. This is consistent with the optical observations in H$\alpha$ and [O III] by Boer et al. (1992), suggesting a young bipolar outflow of age $t \sim 3.5$Myr, that emanates from the nuclear starburst region. Also, Yoshida et al. (1999) proposed a warped disk morphology surrounding the nuclear region, where an inner H I molecular disk is compressed by the bipolar wind, forming a dense ring with active star formation due to gravitational instabilities (Fig. 1).

High-resolution CO observations of NGC 2782, obtained with the IRAM PdBI (now *NOEMA* [1]) interferometer, show emission aligned with the stellar nuclear bar of radius ∼ 1 kpc. The gas traced by CO emission is infalling towards the centre as a result of gravity torques from the nuclear bar (Hunt et al. 2008).

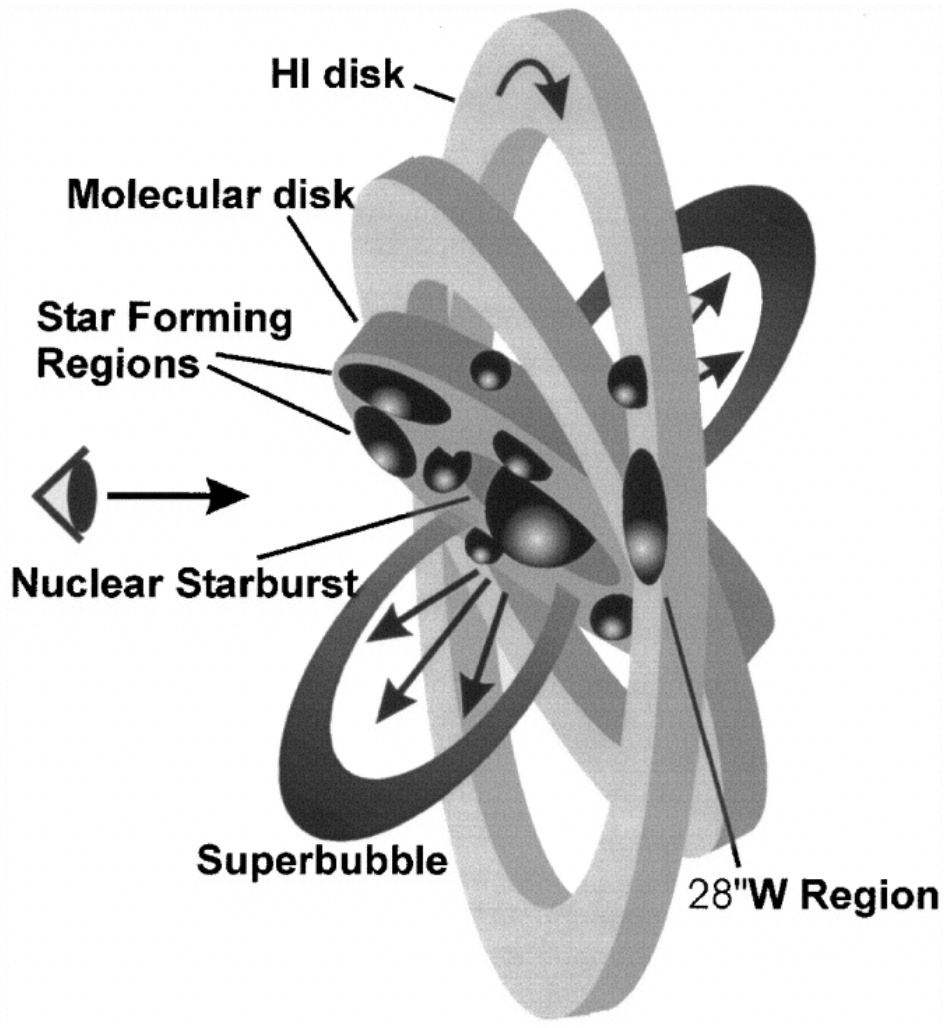

**Figure 1.** Schematic diagram of the structure of the starburst region of NGC 2782 from Yoshida et al. (1999). The nuclear starburst region is surrounded by the highly warped H I molecular disk. The collective effect of supernovae and stellar wind from the nuclear starburst drives the bipolar superbubble out of the inner galactic plane. The inner H I molecular disk is compressed radially by the wind and a dense gaseous ring is formed, then the gravitational instability in the ring leads to active star formation.

The H I tidal tails hosts recent star formation: Several star-forming regions were found in the FUV and NUV by Torres-Flores et al. (2012) in the western tail, as well as in H$\alpha$ (Werk et al. 2011). Knierman et al. (2013) found [C II] emission at the location of the three most luminous H$\alpha$ sources in the eastern tail, but also a lack of CO and [C II] associated with the brighter H$\alpha$ sources in the western tail. They suggested that the western tail material may be undergoing its first star formation and that the star formation efficiency (SFE) is increased by gravitational compression, due to its tidal nature.

In this paper we present results from a ∼ 30 ksec *Chandra* observation of the starburst galaxy NGC 2782. In Section 2, we discuss the *Chandra* observations and data reduction. In Section 3, the analysis of the point sources encountered, their counterparts and the ultraluminous X-ray candidates. In Section 4 we discuss the extended X-ray emission. We present the results of the spectral analysis of the extended emission, the central region and the southern bubble in Section 5, and summarise in Section 6.

## 2 OBSERVATIONS AND DATA REDUCTION

NGC 2782 was observed with the *Chandra* Advanced CCD Imaging Spectrometer *ACIS-S* on May 15, 2002 (ObsID 3014), for 29.96 ksec, with the source located on the *S3* chip. In this paper we only include data from the *S3* chip.

These data were reduced and analysed with the CIAO[2]

[1] Northern Extended Millimetre Array

[2] *Chandra* Interactive Analysis of Observations http://cxc.harvard.edu/ciao

version 4.7, *CALDB* (4.5.6), *HEASoft* (6.12) and *Xspec* (v12.7.0) software packages. The *Chandra* reprocessing script was used to run the data processing threads from CIAO *ACIS* data preparation, set the specific Bad Pixel files, filter lightcurves and to create a new Level 2 Event file. After filtering on good time intervals (GTI) the effective exposure time was 29.58 ksec. The image analysis was restricted to the *ACIS-S3* (*ccd_id* = 7) and the energy range of 0.3–10 keV.

In addition, we also searched the *Galaxy Evolution Explorer (GALEX)*, *Infrared Astronomical Satellite (IRAS)*, *NRAO Very Large Array (VLA)*, *Sloan Digital Sky Survey (SDSS)*, *Hubble Space Telescope (HST)*, and the *Two Micron All-Sky (2MASS)* archives for images and point sources associated with the galaxy.

## 3 POINT SOURCES

### 3.1 X-Ray Point-source detection

We used the CIAO *wavdetect* (Freeman et al. 2002) routine to detect discrete sources, using an exposure map in order to suppress false positives. We choose this script above the other routines (e.g *celldetect* or *vtpdetect*) for its ability to separate closely-spaced point sources and its accurate position refinement algorithm. The exposure map for the *ACIS-S3* chip (*ccd_id* = 7) was obtained by running the *fluximg* script, using an effective energy of 1.5 keV for the broad (0.3–10 keV) band. For the first stage of the routine, we used radii of 2.0 and 4.0 pixels for the wavelet scales (Mexican-Hat wavelet function) and significance threshold of $10^{-6}$. For each X-ray energy bands, the soft (0.3–1.2 keV), the medium (1.2–2.0 keV) and the hard (2.0–10.0 keV) band, we also obtained an exposure map, with assumed effective energies of 0.92 keV, 1.56 keV and 3.8 keV respectively, as recommended by the *fluximg* routine.

We detected a total of 24 sources at the soft band, 28 sources at medium band and 19 sources at the hard band, shown in Fig. 2. This 3-colour image was obtained from using the three bands: soft (red), medium (green) and hard (blue) band images, all with a Gaussian smooth with $\sigma = 8$ pixels.

We then compared our X-ray source lists, identified the position of each source and its re-occurrence in more than one band. After this, we identified a total of 45 point sources on the *ACIS-S3* chip, with positions and properties listed in Table 1.

To estimate the X-ray fluxes listed in Table 1, we used the *srcflux* script from CIAO, using the *wavdetect* source region list and creating a background region for each source in the region list (using the *roi* tool, which also checks for overlapping sources). For the model tool *model flux* run by the *srcflux* script, we used an absorbed power law model with photon index $\Gamma = 1.7$ and Galactic H I column density of $1.76 \times 10^{20}$ cm$^{-2}$ (Dickey & Lockman 1990). Then we obtained the 0.3–10 keV luminosities, which range from 0.14 to 40.9 $\times 10^{39}$ erg s$^{-1}$, assuming a distance to NGC 2782 of 37 Mpc (Smith et al. 2012). Some of the background regions obtained using the *roi* tool had very low counts rates. When the background region had $< 2$ counts, we changed the background region and obtained the spectra separately and fitted it using a power law model with the same specifications previously described.

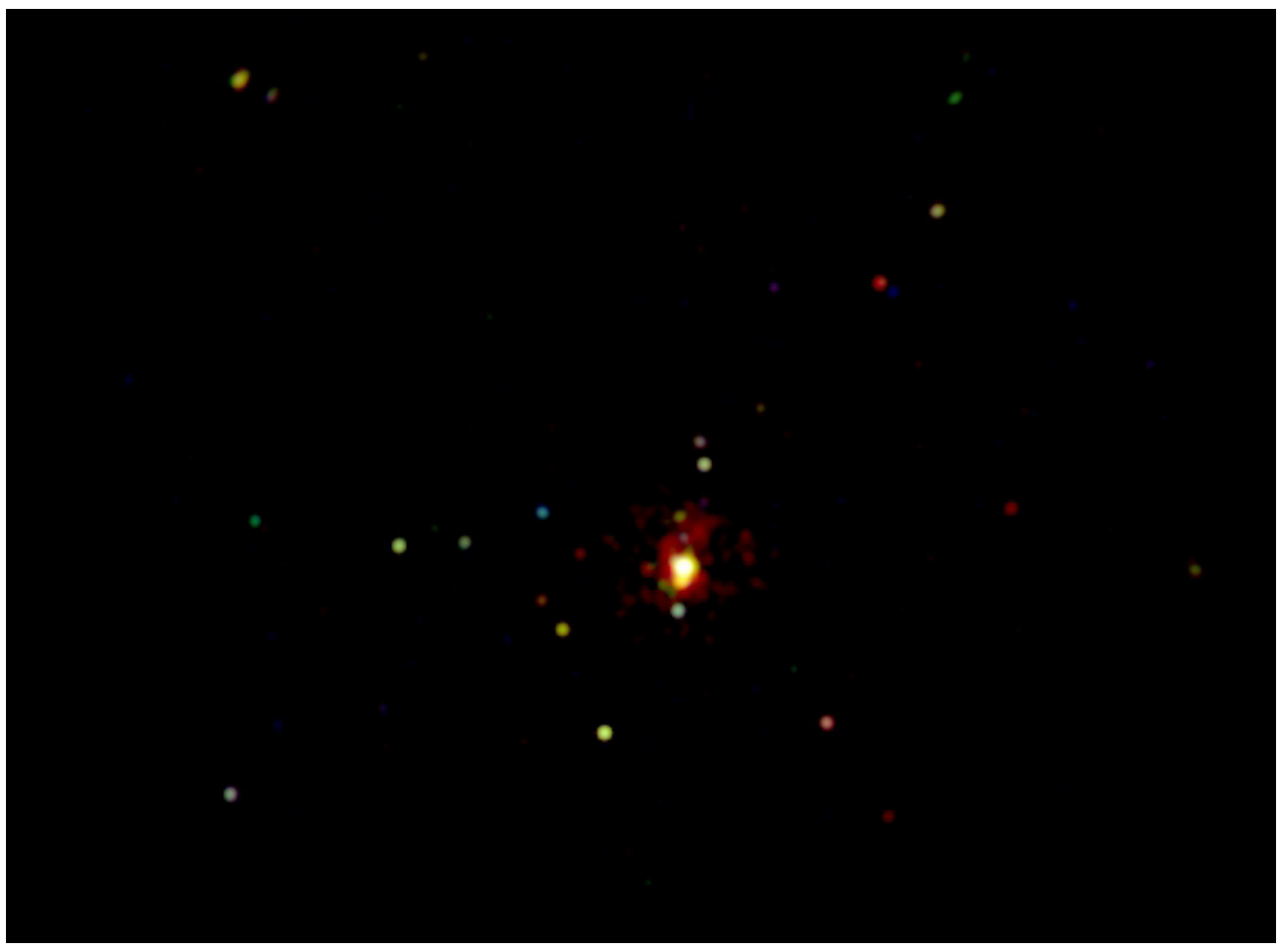

**Figure 2.** Three colour *Chandra ACIS* image of NGC 2782, showing diffuse emission and 45 point sources. The energy band associated with the colours are: red for 0.3–1.2 keV; green for 1.2–2.0 keV; blue for 2.0–10.0 keV. The size of the image is $\sim 8 \times 8$ arcmin. North is up, East is left.

In Fig. 3 we show the DSS image of NGC 2782, along with the $D_{25}$ isophotal diameter [3](Makarov et al. 2014). The $D_{25}$ ellipse region has a major axis of 1.62 arcmin and minor axis of 1.35 arcmin, at an angle of 20 degrees. In this figure we also show the X-ray contours, with the extended X-ray emission confined to the central regions of the galaxy.

A density plot of the point sources is shown in Fig. 4, with radii increments of 0.5 arcmin, showing the $D_{25}$ major axis (1.62 arcmin; dotted line). We used the result of Kim et al. (2007) to calculate the background source fraction expected for the broad energy band. If we take a limit of source flux of $\sim 10^{-15}$ erg cm$^{-2}$ s$^{-1}$, which corresponds to $\sim$ 25 cts, then we expect around 7 background sources per chip.

We cross-checked our source list with that from the *Chandra Source Catalog* [4] (Evans et al. 2010), linked with the *ACIS-S3* chip and obtained from the same observation (ObsID 3014). The CSC found 41 sources in the same region as considered here, all included in our source list, except for two additional point sources not found by our source extraction method. The CSC catalogue listed these unresolved sources to be very near the central source, in a region which appeared as diffuse emission in the images, and will be discussed in Section 5.

### 3.2 Counterparts

In order to identify counterparts to our X-ray sources, we searched the point source catalogues for optical (SDSS, Release 9) and IR (2MASS) counterparts and found several point sources with spatial correspondence (Table 1). We found 21 optical (SDSS) and 5 IR (2MASS) matches, apart from the central source, within a search radius of 8 arcmin and a position error of 5 arcsec.

Some objects may be associated with the tidal regions of the galaxy, located near star formation zones and others with high luminosity may be background or foreground objects. Optical spectroscopic follow-up will be difficult. At a distance of 37 Mpc, a B0V star would have an apparent magnitude of +28.7 and an O5 supergiant ($M_V = -6.4$) would have $V = 26.4$ (Zombeck 1990). On the same grounds, Gladstone et al. (2013) suggested a distance limit of $\lesssim 5$ Mpc in order to detect optical counterparts of X-ray sources.

[3] http://leda.univ-lyon1.fr/
[4] http://cxc.harvard.edu/csc/

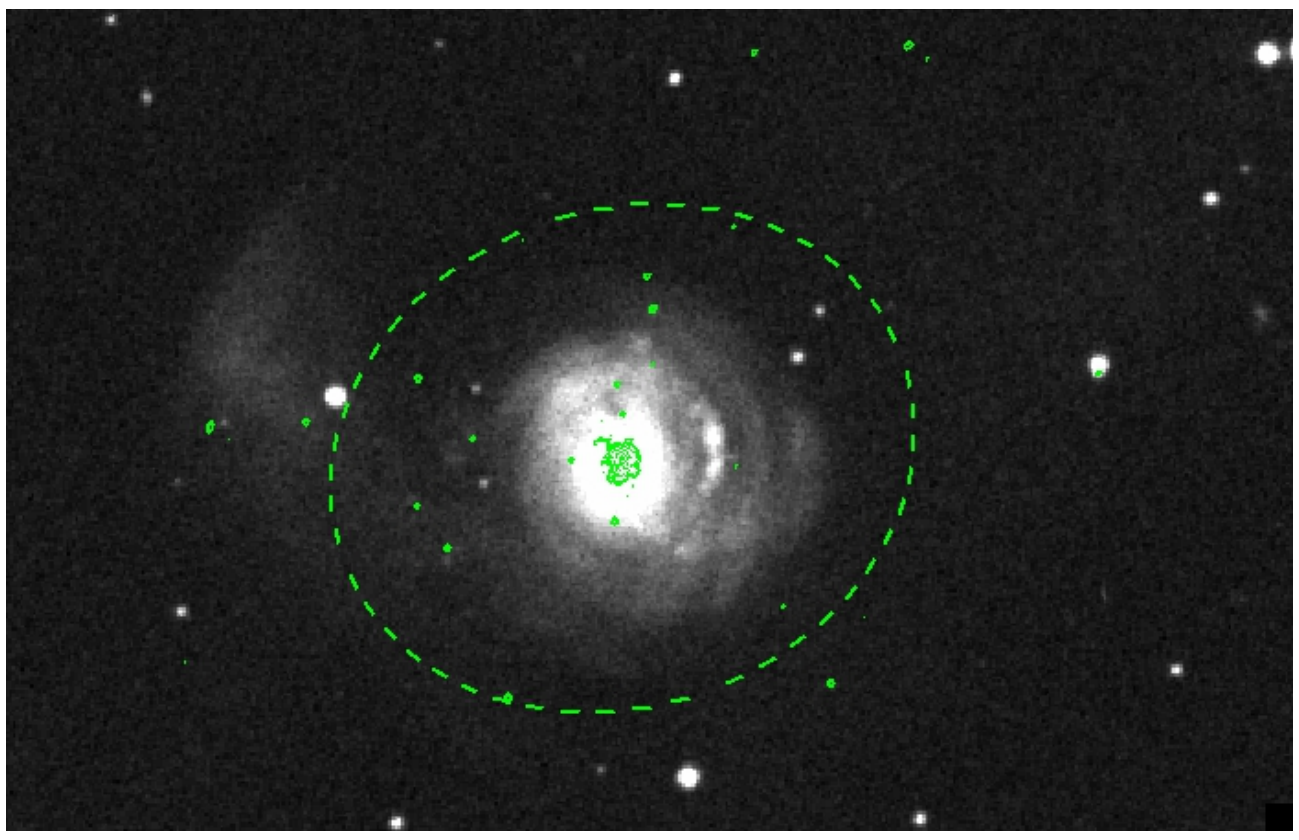

**Figure 3.** $D_{25}$ ellipse of 1.62 arcmin major radius, minor radius of 1.35 arcmin at 20°, shown in an optical image of NGC 2782. X-ray contours in green.

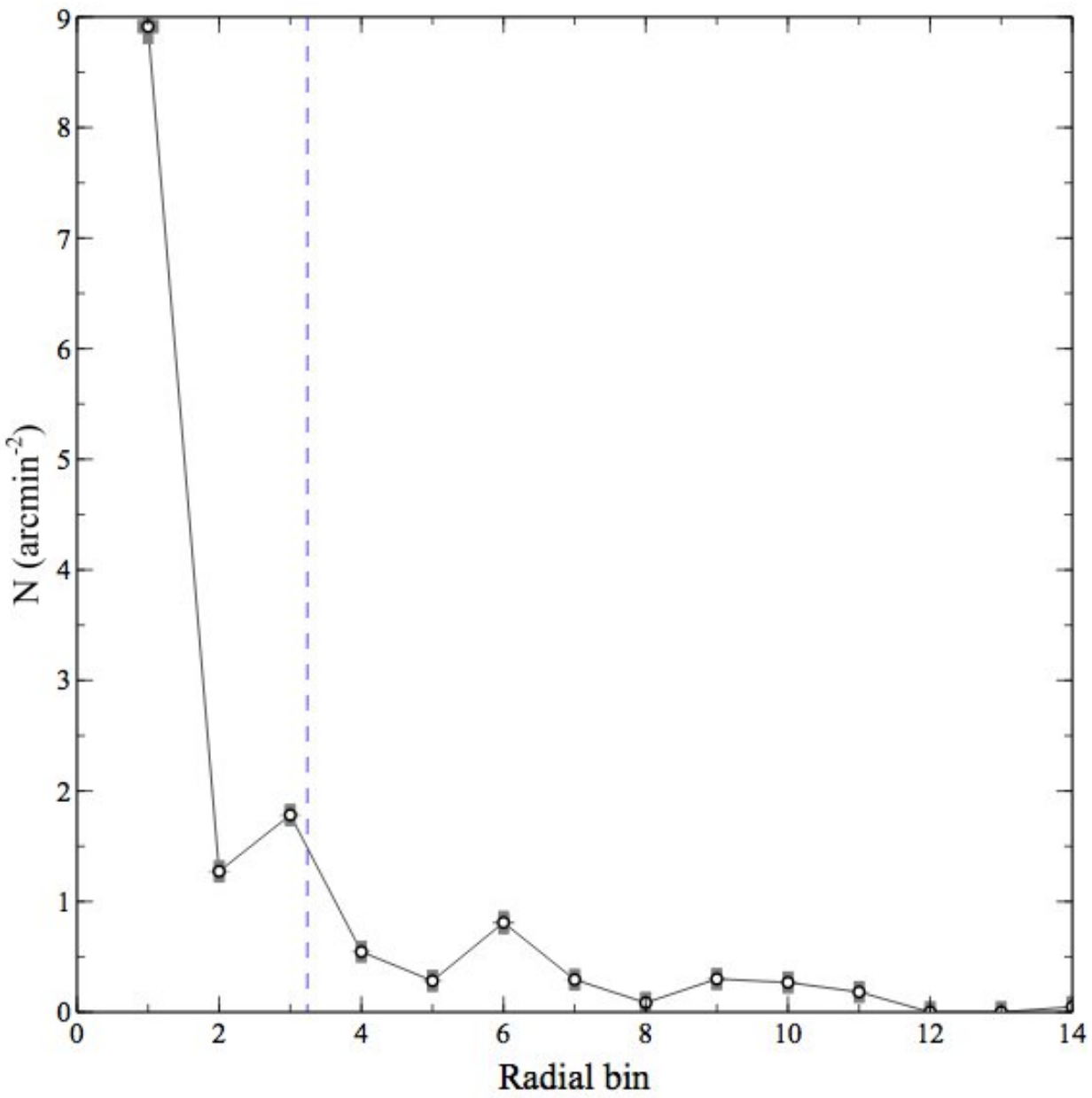

**Figure 4.** Number of sources per square arcmin vs. radial bin. Each radial bin has a 0.5 arcmin difference and the dotted line shows the $D_{25}$ radial limit of 1.62 arcmin. The background level is 0.11 src arcmin$^{-2}$.

Figure 5 (left) is a mosaic image of NGC 2782 showing the optical, H I and FUV features. Given the extent of the tidal tails, we looked for possible point sources and counterparts correlated to the galaxy and, with this in mind, we choose to include point sources that are within a region of $2D_{25}$ radius (Table 2) of the galaxy centre. A total of 27 sources fit that criteria, 14 of them are ULXs (including the central source for the time being), 16 of them have an optical counterpart and 4 of them an IR counterpart. The IDs of the individual sources can be identified in Fig.5 (right). We classified the sources by their apparent location within the galaxy (i.e. nuclear, disk, external disk) and we list the SDSS object type in Table 2. Considering the area of the $2D_{25}$ region and background number of sources per square degree (see Section 3.1), we expect $\sim 3.6$ background sources in the total area and 2.7 sources in the annulus region between a radius of $D_{25}$ and $2D_{25}$. This is for detectable X-ray sources, for ULXs the expected number of background objects is $< 1$.

We have plotted the X-ray luminosity function (XLF) of these 27 point sources (i.e. within $2D_{25}$; Fig. 6). Fitting the XLF using a simple power-law model, the best fit produced a slope of $\gamma = -0.63 \pm 0.16$, close to the typical value of $\sim -0.6$ found in the luminosity distributions of HMXB (Grimm et al. 2003, Swartz et al. 2011, Mineo et al. 2012). We also used a power-law model with an exponential cut-off to fit the XLF and for this model found a cut-off luminosity of $L_{cut} = 1.5 \times 10^{40}$ erg s$^{-1}$.

### 3.3 Ultraluminous X-Ray Point Sources

Ultraluminous X-ray point sources (ULXs), defined as those with an intrinsic luminosity of $L_X \geqslant 10^{39}$ erg s$^{-1}$ (Gladstone et al. 2009; Swartz et al. 2004), are common in starburst galaxies. We also require the ULXs to be non-nuclear and to be compact (i.e. not extended).

Studies of ULXs in spiral galaxies have shown a correlation between the number of ULXs and the galaxy star formation rate, with some of these studies suggesting that they are associated with high-mass X-ray binaries (HMXBs; Swartz et al. 2004, Swartz et al. 2011) and that they are a good tracer of recent star-formation activity (Grimm et al. 2003, Mineo et al. 2012), being also consistent with the relation between the hard (2–10 keV) X-ray luminosity and the SFR (Ranalli et al. 2003, Persic & Rephaeli 2007). There is also evidence that suggests that ULXs are more common in low metallicity galaxies (Mapelli et al. 2010; Mapelli et al. 2011; Kaaret et al. 2011;Prestwich et al. 2013 ).

For our detected sources, we used an absorbed power law model with photon index $\Gamma = 1.7$ and a Galactic H II column density of $1.76 \times 10^{20}$ cm$^{-2}$ to find fluxes and luminosities (Table 1). As discussed above, we identify 13 non-nuclear sources as being ULXs associated with NGC 2782, with X-ray luminosities of $1.2 - 3.9 \times 10^{39}$ erg s$^{-1}$.

Our ULXs sample is similar to sources listed by Smith et al. (2012), with the exception of 2 non-resolved point sources in the central region, that were not detected by the *wavdetect* routine and we consider them to be embedded into the diffuse emission.

With 13 ULXs with $L_x > 10^{39}$ erg s$^{-1}$, NGC 2782 seems to have a surprisingly large population of ULXs. Using results from Swartz et al. (2004) and Swartz et al. (2011) for the $N(ULX)/L_B$ ratio for their sample of spiral galaxies, the expected number of ULXs in NGC 2782 with an X-ray luminosity $L_x > 10^{39}$ erg s$^{-1}$ (0.3 − 8.0 keV) is $\sim 3 - 6$, assuming an absolute B-band magnitude of −20.94.

The Swartz et al. (2004) sample also shows a correlation between $N$(ULX) and the far-infrared luminosity

**Table 1.** Complete *ACIS-S3* list of detected point sources and properties

| Source ID | R.A. (J2000) | Decl. (J2000) | NET_COUNTS counts | Flux [a] $10^{-14}$ [erg cm$^{-2}$ s$^{-1}$] | $L_X$ [b] $10^{39}$ [erg s$^{-1}$] | COUNTERPART [c] | LOCATION[d] |
|---|---|---|---|---|---|---|---|
| 1 | 09 14 23.82 | +40 05 01.7 | 55.82 | 2.15 | 3.50 | Optical | Exterior |
| 2 | 09 14 08.38 | +40 05 30.8 | 105.57 | 2.32 | 3.80 | Optical | Disk |
| 3 | 09 13 59.19 | +40 05 35.6 | 53.86 | 1.46 | 2.39 | Optical | Extended Disk |
| 4 | 09 14 17.57 | +40 05 42.6 | 9.59 | 0.44 | 0.71 | | Extended Disk |
| 5 | 09 14 00.55 | +40 06 00.9 | 9.00 | 0.30 | 0.50 | | Disk |
| 6 | 09 14 10.11 | +40 06 19.7 | 43.85 | 0.70 | 1.15 | Optical | Disk |
| 7 | 09 14 05.34 | +40 06 28.7 | 63.58 | 1.83 | 2.99 | | Disk |
| 8 | 09 14 10.97 | +40 06 33.5 | 15.00 | 0.24 | 0.40 | Optical | Disk |
| 9 | 09 14 05.20 | +40 06 42.7 | 185.00 | 2.37 | 3.87 | | Disk |
| 10 | 09 14 06.57 | +40 06 48.6 | 15.58 | 0.18 | 0.30 | Optical | Disk |
| 11 | 09 14 05.11 | +40 06 48.9 | 647.27 | 24.97 | 40.9 | Optical,IR [d] | Nuclear |
| 12 | 09 14 05.43 | +40 06 51.7 | 76.09 | 1.66 | 2.72 | | Disk |
| 13 | 09 14 09.38 | +40 06 55.6 | 12.86 | 0.35 | 0.57 | Optical, IR | Disk |
| 14 | 09 14 16.87 | +40 06 59.4 | 65.72 | 1.71 | 2.80 | | Extended Disk |
| 15 | 09 14 14.14 | +40 07 01.2 | 29.18 | 0.76 | 1.25 | Optical | Extended Disk |
| 16 | 09 14 05.08 | +40 07 03.3 | 16.57 | 0.50 | 0.82 | | Disk |
| 17 | 09 14 05.26 | +40 07 13.2 | 20.08 | 0.28 | 0.47 | Optical | Disk |
| 18 | 09 14 10.94 | +40 07 15.4 | 35.59 | 1.63 | 2.67 | | Extended Disk |
| 19 | 09 13 51.56 | +40 07 17.2 | 13.46 | 0.18 | 0.30 | Optical, IR | Extended Disk |
| 20 | 09 14 04.25 | +40 07 20.1 | 10.72 | 0.30 | 0.48 | | Disk |
| 21 | 09 14 04.25 | +40 07 38.1 | 64.44 | 1.56 | 2.56 | Optical | Disk |
| 22 | 09 14 04.43 | +40 07 48.7 | 29.00 | 0.80 | 1.31 | | Extended Disk |
| 23 | 09 14 01.93 | +40 08 04.8 | 10.72 | 0.16 | 0.26 | | Disk |
| 24 | 09 13 56.43 | +40 08 59.9 | 8.86 | 0.32 | 0.20 | Optical | Extended Disk |
| 25 | 09 14 01.36 | +40 09 02.0 | 16.00 | 0.93 | 1.52 | Optical | Extended Disk |
| 26 | 09 13 56.97 | +40 09 04.1 | 38.86 | 0.79 | 1.30 | Optical | Extended Disk |
| 27 | 09 13 54.62 | +40 09 38.1 | 49.34 | 1.46 | 2.39 | Optical | Exterior |
| 28 | 09 13 49.91 | +40 09 52.7 | 2.94 | 0.12 | 0.19 | | Exterior |
| 29 | 09 14 23.47 | +40 10 40.3 | 81.82 | 1.29 | 2.11 | Optical | Exterior |
| 30 | 09 14 06.57 | +40 04 19.8 | 6.00 | 0.22 | 0.36 | | Exterior |
| 31 | 09 13 56.68 | +40 04 51.0 | 6.85 | 0.08 | 0.14 | Optical, IR | Extended Disk |
| 32 | 09 13 43.98 | +40 06 48.7 | 7.46 | 0.31 | 0.50 | Optical, IR | Exterior |
| 33 | 09 14 15.38 | +40 07 08.1 | 4.85 | 0.14 | 0.23 | | Extended Disk |
| 34 | 09 14 22.80 | +40 07 11.4 | 18.21 | 1.07 | 1.75 | Optical | Exterior |
| 35 | 09 13 45.84 | +40 08 25.2 | 8.81 | 0.86 | 1.41 | | Exterior |
| 36 | 09 14 05.12 | +40 09 30.2 | 7.72 | 0.41 | 0.67 | Optical | Extended Disk |
| 37 | 09 14 04.46 | +40 10 10.8 | 5.86 | 1.19 | 1.94 | | Exterior |
| 38 | 09 13 53.87 | +40 10 31.5 | 21.41 | 1.34 | 2.19 | | Exterior |
| 39 | 09 14 22.14 | +40 10 33.0 | 28.82 | 1.12 | 1.84 | | Exterior |
| 40 | 09 14 15.89 | +40 10 51.0 | 11.58 | 0.25 | 0.40 | | Exterior |
| 41 | 09 13 55.47 | +40 11 16.0 | 10.17 | 1.33 | 2.18 | | Exterior |
| 42 | 09 14 19.92 | +40 11 18.9 | 29.68 | 1.35 | 2.20 | IR | Exterior |
| 43 | 09 13 56.97 | +40 11 20.8 | 7.48 | 0.62 | 1.02 | | Exterior |
| 44 | 09 14 08.67 | +40 11 59.9 | 55.89 | 3.17 | 5.19 | | Exterior |
| 45 | 09 13 52.41 | +40 13 12.6 | 110.85 | 4.63 | 7.58 | Optical | Exterior |

Notes:
[a] Assuming an absorbed power law with photon index $\Gamma = 1.7$ and Galactic H I column density of $1.76 \times 10^{20}$ cm$^{-2}$ (Dickey & Lockman 1990)
[b] For 0.3–8.0 keV, assuming a distance of 37 Mpc
[c] Point source catalogues for optical (*SDSS* Release 9) and Infrared (*2MASS*) counterparts
[d] Based on the apparent position from the nucleus of the galaxy and relative to the $D_{25}$ isophotal diameter being nuclear the central point source, disk the extended region $< D_{25}$, the extended disk de annulus region $D_{25} < d < 2D_{25}$ and exterior region $> D_{25}$.
[e] IRAS

of the sample spiral galaxies, with $N(\mathrm{ULX}) = (0.022 \pm 0.01)L_{FIR}/(10^{42}\,\mathrm{erg\ s^{-1}}) + (0.64 \pm 0.3)$. From this, and assuming an infrared luminosity of $L_{FIR} = 10^{10.31} L_{\odot}$ for NGC 2782, the expected number of ULXs is $\sim 2.5$.

On the other hand, the number of ULXs candidates at a luminosity of $L_x \geqslant 10^{40}$ erg s$^{-1}$ is congruent with the expected value of $< 1$, using the $N(\mathrm{ULX})/L_B$ ratio found by Smith et al. (2012), for their Arp interacting galaxy sample.

Prestwich et al. (2013) considered the number of ULX per unit star-formation rate ($N_{ULX}(SFR)$). They found the highest values of $N_{ULX}(SFR)$ for those galaxies with $12 + \log(O/H) < 7.65$, while for the high metallicity sample ($12 + \log(O/H) > 8.0$) they found no significant trend with metallicity. For the low metallicity sample the typical value of $N_{ULX}(SFR)$ was around 7 per $M_\odot$ yr$^{-1}$ of SFR (though with large error bars), while for the high metallicity sample, the values of $N_{ULX}(SFR)$ were much lower (mean value 0.17).

For the observations of HII regions by Werk et al. (2011), NGC 2782 has a metallicity of $12+\log(O/H) \sim 8.5$, meaning NGC 2782 is a high metallicity galaxy in the context of Prestwich et al. (2013).

For NGC 2782 we estimate the SFR as 6.27 $M_\odot$ yr$^{-1}$ (see Section 5). For NGC 2782 we have a total of 13 ULX sources, of which $< 1$ should be a background source. In this case, then $N_{ULX}(SFR) \sim 2$. This means NGC 2782 lies well above the other galaxies in Prestwich et al. (2013) and does seem to have a larger number of ULX sources than might be expected from its SFR.

Altogether, this suggests NGC 2782 has an unusually large number of ULX sources, though the relatively small number of sources means it is not hugely significant.

**Table 2.** Point Sources within 2 $D_{25}$ ellipse. Sources above the dotted line are ULXs.

| Source ID [a] | $L_X$ [b] $10^{39}$ | Location [c] | Counterpart [d] | SDSS Type [e] |
|---|---|---|---|---|
| 11 | 40.9 | Nuclear | Optical, IR | 3 |
| 9 | 3.9 | Disk | | |
| 2 | 3.8 | Disk | Optical | 6 |
| 7 | 3.0 | Disk | | |
| 14 | 2.8 | Extended Disk | | |
| 12 | 2.7 | Disk | | |
| 18 | 2.7 | Extended Disk | | |
| 21 | 2.6 | Disk | Optical | 6 |
| 3 | 2.4 | Extended Disk | Optical | 6 |
| 25 | 1.5 | Extended Disk | Optical | 6 |
| 22 | 1.3 | Extended Disk | | |
| 26 | 1.3 | Extended Disk | Optical | 6 |
| 15 | 1.3 | Extended Disk | Optical | 3 |
| 6 | 1.2 | Disk | Optical | 6 |
| 4 | 0.7 | Disk | | |
| 36 | 0.7 | Extended Disk | Optical | 3 |
| 13 | 0.6 | Disk | Optical, IR | 6 |
| 17 | 0.5 | Disk | Optical | 6 |
| 8 | 0.4 | Disk | Optical | 6 |
| 30 | 0.4 | Extended Disk | | |
| 10 | 0.3 | Disk | Optical | 6 |
| 19 | 0.3 | Disk | Optical, IR | 6 |
| 23 | 0.3 | Disk | | |
| 33 | 0.2 | Disk | | |
| 24 | 0.2 | Extended Disk | Optical | 3 |
| 31 | 0.1 | Extended Disk | Optical, IR | 6 |
| 5 | 0.1 | Disk | | |

Notes:
[a] Source ID from Table 1
[b] In erg s$^{-1}$, assuming an absorbed power law with photon index $\Gamma = 1.7$, Galactic H I column density of $1.76 \times 10^{20}$ cm$^{-2}$ (Dickey & Lockman 1990), energy range of 0.3–8.0 keV, assuming a distance of 37 Mpc
[c] Based on the apparent position from the nucleus of the galaxy and relative to the $D_{25}$ isophotal diameter being nuclear the central region, disk the extended region $< D_{25}$, the extended disk de annulus region $D_{25} < d < 2D_{25}$ and exterior region $> D_{25}$.
[d] Point source catalogues for optical (*SDSS* Release 9) and Infrared (*2MASS*) counterparts
[e] SDSS Type: 3=Galaxy, 6=Star.

## 4 IMAGING ANALYSIS

### 4.1 X-ray Morphology and features

In Fig. 7 and Fig, 8 we show the X-ray morphology of the central regions of NGC 2782. Fig. 7 shows a broad band (0.3–10 keV) image, which has been Gaussian smoothed with $\sigma =$ 3 pixels. From this image we see elongated diffuse emission in the central regions of NGC 2782, with a total size scale of $\sim 0.5$ arcminute North to South, and $\sim 0.25$ arcmin East to West.

We also show in Fig. 7 two surface brightness profiles, along the major and minor axes (units of counts per pixel, with the slices having a width of 6 pixels). In this plot the typical $\sigma$ would be approximately 0.4 counts per bin. The 'major axis profile is taken from North to South, at an angle of 250 °(angle increasing anticlockwise from the horizontal), chosen to be coincident with the Saikia et al. (1994) 5 GHz radio bipolar emission. The 'minor axis' profile from East to West, is at an angle of 340° (with the angle defined as above), perpendicular to the North to South projection. From this we can determine that the projected physical extent of the diffuse X-ray emission is 5.4 kpc by 2.7 kpc.

From Fig. 7, we see extended emission features that likely indicates the presence of a young galactic superwind. In particular, a shell-like structure located $\sim 7$ arcsec south of the core source is detected in X-rays, and also H$\alpha$ and radio images (see Section 4.3). This structure potentially extends to $\sim 1.9$ kpc, considering the face-on geometry of the galaxy with inclination angle of $i = 47°$ (Boer et al. 1992). The absence of a corresponding northern bubble can be interpreted either as an intrinsic absence of X-rays (perhaps due to asymmetries in the starburst or the interstellar medium), or because of heavy absorption, obscuring the emission.

In Fig. 8 we show the diffuse emission in a three colour image in the soft, medium and high energy bands. This image is coded with red for the 0.2 –1.5 keV band ; green for the 1.5–2.5 keV band; and blue for the 2.5–8.0 keV band, respectively. For this image the data has been adaptively smoothed using the *dmimgadapt* script from CIAO, using a top-hat function with radii/scales ranging from 0.1 to 30 pixels, with 50 scales spaced logarithmically between the minimum and maximum size scales.

### 4.2 Multiwavelength image analysis

We have compared the X-ray morphology with several other bands, with data taken from the respective archives. In particular we compare with H$\alpha$ emission (which shows ionised gas associated with hot stars or an outflow), FUV emission (associated with the presence of hot stars and massive star-formation), 5 GHz radio emission (associated with syn-chrotron emission) and 1.4 GHz emission (associated with neutral gas and tracing out evidence of ongoing merging).

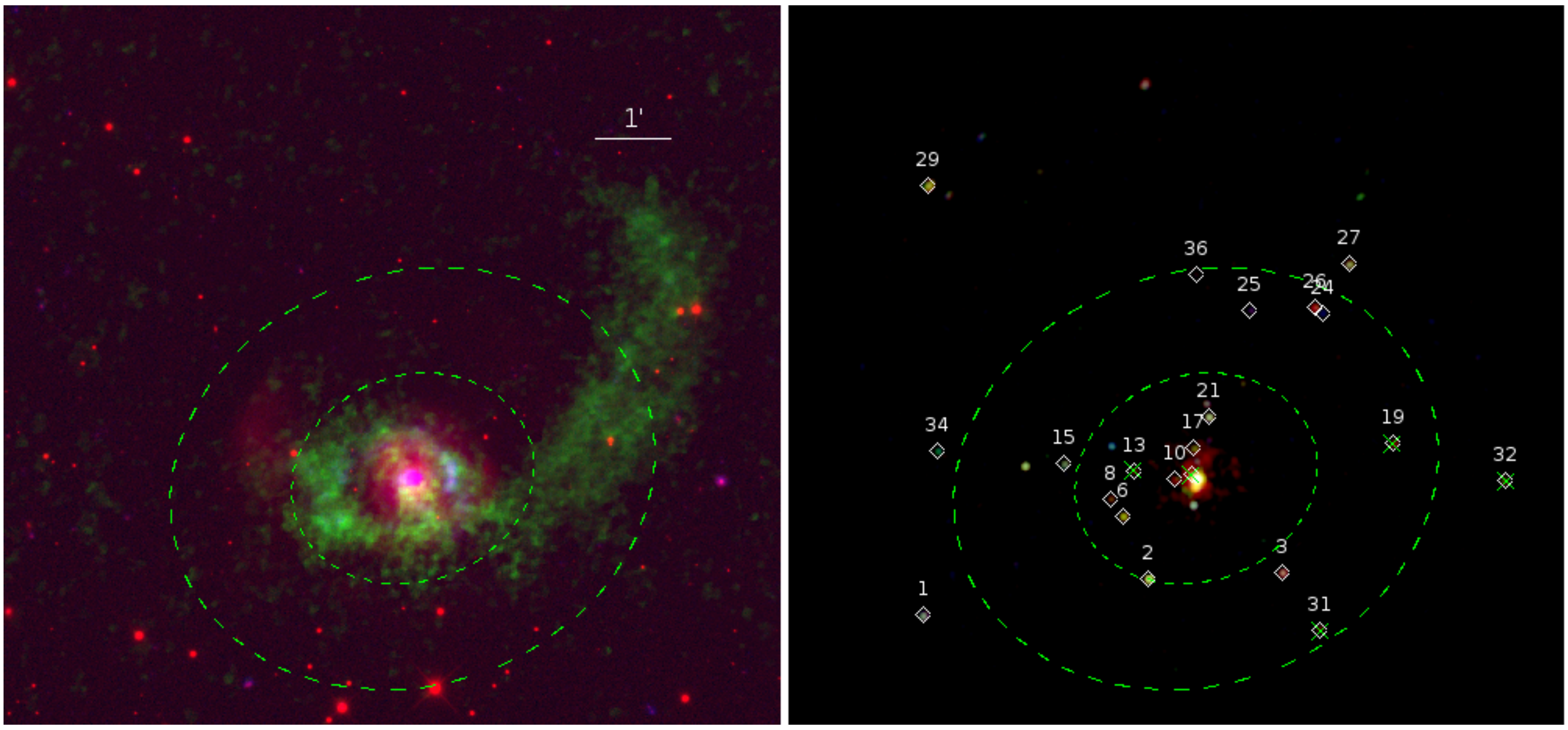

**Figure 5.** Extended region of 2 $D_{25}$ radius of NGC 2782 at the $\sim 8 \times 8$ arcmin field of view. *Left:* RGB montage of the galaxy, showing its morphology. Red corresponds to red filter SDSS image, green to 21 cm image showing the extent of the tidal H I tail (Smith 2004?) and blue to GALEX FUV image. The western tail extends beyond 2 times the isophotal diameter. *Right:* Optical (SDSS release 9) and IR (2MASS) counterparts on a *ACIS-S3* (*ccd_id* = 7) three-color image, with energy range of 0.3–8.0 keV . White diamonds refer to X-ray sources matching the position of optical sources, green crosses indicate the sources with optical and IR counterparts, both with a position error of 5 arsec.

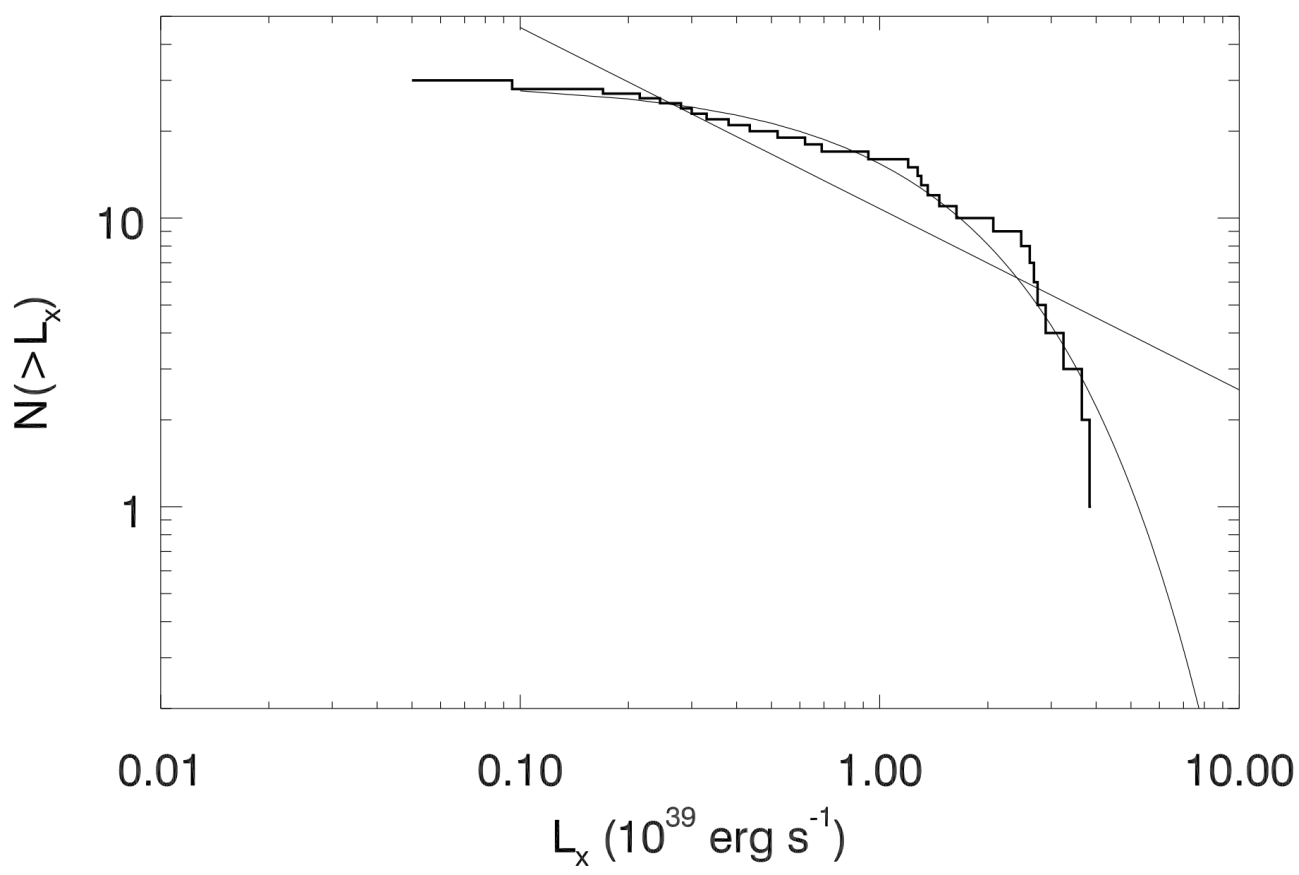

**Figure 6.** X-ray luminosity distribution of the point sources detected within the 2 $D_{25}$ ($\sim 3.24$ arcmin) ellipse. The best-fitting power-law model with $\gamma = -0.63 \pm 0.16$ is shown by the green dashed line. A power-law model with an exponential cut-off is also shown.

Analysing multiwavelength images of the galaxy, we have found clear correspondence between aspects of the X-ray and optical, radio and UV emission. The X-ray emission is only located in the central regions of the galaxy (see Fig.9), where the X-ray contours are superimposed on the optical and GALEX FUV images of NGC 2782.

Fig.10 shows the X-ray data (presented as contours) superimposed on the HST H$\alpha$ emission (data taken on April 18, 1997, with the WFPC2, P.I. Baum). Perhaps most interestingly, the southern X-ray feature, spatially corresponds with the filaments of the H$\alpha$ southern emission, which has been interpreted as being associated with an outflow (Boer et al. 1992). In several starburst galaxies (for instance NGC 1482, Strickland et al. 2004a and Vagshette et al. 2012 ; NGC 3079, Cecil et al. 2002, Strickland et al. 2003 and Strickland et al. 2004a; M82, Stevens et al. 2003) the general morphology of the diffuse X-ray emission is similar to the filamentary morphology of the H$\alpha$ emission, and the line-emitting features are often limb-brightened, indicating that much of the optically emitting gas resides on the surface of hollow structures. The soft X-ray emission near the filaments suggests a partially filled volume of warm gas which has been swept up and shock-heated by the superwind (Strickland & Stevens 2000, Strickland 2002, Strickland et al. 2004a,b, Veilleux et al. 2005).

Several starburst galaxies have large-scale non-thermal radio halos. These expanding structures arise as relativistic magnetised plasma is moved by the superwind out of the central starburst. Studies of starburst galaxies with halos of synchrotron emission (M82, Seaquist & Odegard 1991; NGC 253, Carilli et al. 1992) have shown a close connection between the radio halo and the system of faint H$\alpha$ filaments extending many kpc out of the galactic disk, and we see a similar situation in NGC 2782, but on a rather smaller scale.

In NGC 2782, *VLA* 5 GHz images shows emission of a bipolar flow from north to south of the galaxy (see Fig. 11). This structure shows considerable overlap with the X-ray and H$\alpha$ emission in the southern part of the outflow (especially with the X-ray feature located 7 arcsec south of the nucleus).

However, we do see an extension in the radio emission to the North of the nucleus. The absence of an obvious northern feature in the X-ray data can be interpreted either as due to a lack of X-rays or heavy absorption (especially given the orientation of the galaxy, see Fig. 1).

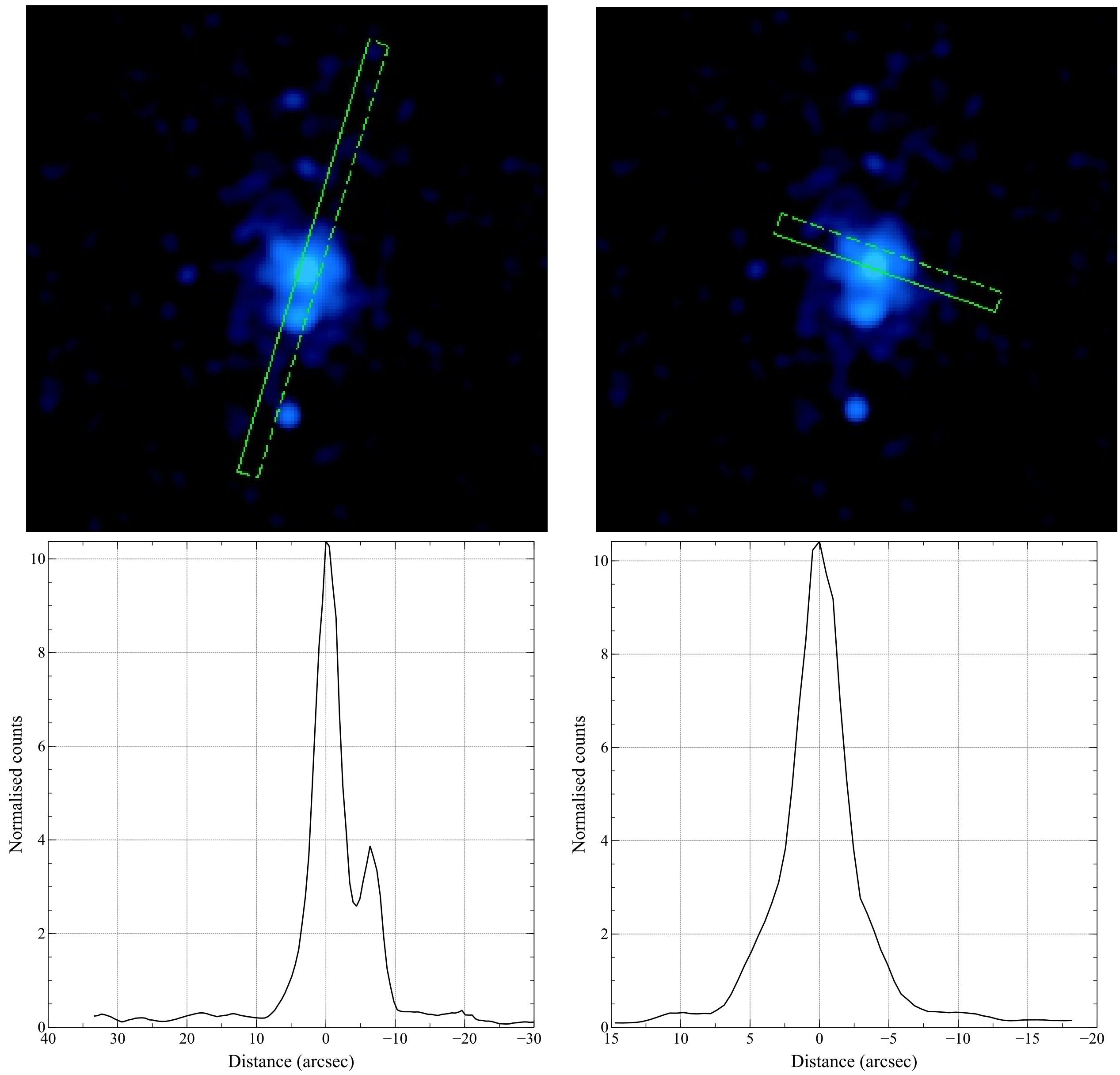

**Figure 7.** 3-arcsec width Projections for the extended emission of NGC 2782. *Left:* North–South 1.06-arcmin length projection of the X-ray emission, at an angle of 250 °(angle increasing anticlockwise from the horizontal), showing extended emission and count peaks to the south of the galaxy. *Right:* East–West projection of the emission, taken from a region of 0.56 arcmin length and 340°(angle define as above). Both plots show average counts vs. position from the nuclear region.

Close consideration of Fig. 10 does give some indication of an X-ray extension (and some H$\alpha$ emission) in the same general location as the northern spur of the radio emission. While we have clear evidence of an outflow in the South, we do have some evidence of a bipolar wind.

Comparing these results with observations of the superwind galaxy NGC 1482, the physical correspondence between the morphologies of the dust, the H$\alpha$ emission lines and the diffuse X-ray emission is similar to what we see in NGC 2782 (Vagshette et al. 2012). However in NGC 1482, the superwind appears to be more extended. In NGC 1482 there is no evidence of a large-scale outflow in the radio continuum emission (Hota & Saikia 2005), and the emission is confined to the disk of the galaxy.

Studies of the interaction of superwinds with cool atomic and molecular gas have shown that superwinds in starburst galaxies immersed in low density intergalactic medium would quickly adiabatically cool, their density would rapidly drop, and they would then become undetectable in X-rays or optical line emission (Heckman et al. 1993). In NGC 2782 we are dealing with a younger developing superwind and more detailed observations in radio

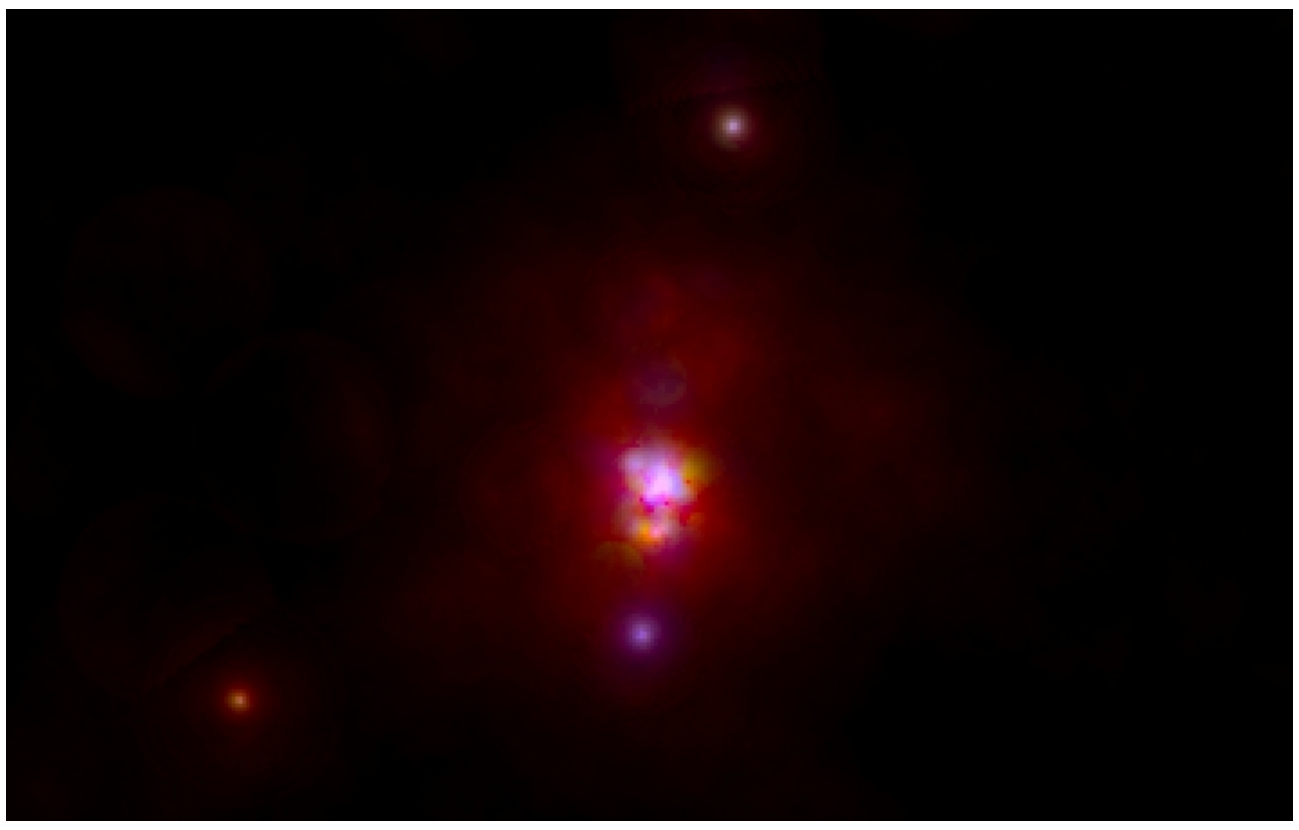

**Figure 8.** Adaptively smoothed three colour *Chandra ACIS* image of NGC 2782, showing diffuse emission extending to $\sim$ 30 arcsec North to South. The energy band associated with the colours are: red for 0.2–1.5 keV; green for 1.5–2.5 keV; blue for 2.5–8.0 keV.

and H$\alpha$ may provide insights into the development of superwinds.

Several lines of evidence suggest that superwinds contain cool, dust-bearing gas (Heckman 2000) and as dust is highly reflective in the ultraviolet, starburst superwinds imaging in this wavelength can trace the location of dust, if one can account for UV emission by photoionised or shock-heated gas. For instance, Hoopes & et al. (2005) established a close morphological correspondence between the dust and the hotter phases of the winds probed in H$\alpha$ and X-ray emission, for NGC 253 and M82. Comparing X-ray emission with the FUV GALEX data of NGC 2782, there is spatial correspondence between them in the central region of the galaxy, as shown in Fig.9. There is no X-ray emission associated the shell-like structure seen in the FUV.

## 5 SPECTRAL ANALYSIS

Spectral analysis from previous X-ray observations of typical starburst galaxies have shown that multiple spectral components are needed to fit the spectrum (e.g. Moran & Lehnert 1997; Ptak et al. 1997; Cappi et al. 1999; Lira et al. 2002; Zhang et al. 2006; Bauer et al. 2008).

For NGC 2782, we have extracted spectra from several regions: i) a central region (associated with a potential low luminosity AGN), ii) the whole region of extended emission and iii) the discrete region of X-ray emission located to the south of the nucleus. In Table 3 we show the results of the best-fitting models for these regions and the results are described below.

### 5.1 Extended emission

In Fig. 12 we show the regions used to generate the spectra and the spectral fits. For the whole extended X-ray emission, the region considered has a radius of 9.8 arcsec, but the central nuclear region has been excluded (with the excluded region having a radius of 1.5 arcsec).

We fitted the spectra with a power law plus a collisionally-ionised diffuse gas (APEC), obtaining the parameters listed on Table 3. We obtained solar abundances, $\Gamma = 1.5$ and temperature of $kT = 0.77 \pm 0.09$ keV, with a reduced $\chi^2_\nu = 1.43$ for 73 degrees of freedom. The spectra shows thermal emission below 2 keV and flat power law on the higher energy emission, typical of a star-forming galaxy.

Apart from the linear relation between radio and far-IR luminosities and the Star Formation Rate (*SFR*), it has been suggested that the hard (2–10 keV) X-ray emission is directly related to the *SFR* as well (Ranalli et al. 2003, Persic & Rephaeli 2007). We used the Ranalli et al. (2003) hard X-ray emission (2–10 keV) and SFR relation to obtain a *SFR* of 2.5 $M_\odot\ yr^{-1}$, from the obtained hard luminosity of $L_X = 1.26 \times 10^{40}$ erg s$^{-1}$. Using the recent linear relation between X-ray emission in the $0.5 - 8$ keV band and SFR from Mineo et al. (2014), we obtained a SFR of 6.75 $M_\odot\ yr^{-1}$, which is just above the estimated SFR of 3-6 $M_\odot\ yr^{-1}$ and also consistent with the the SFR relation of Kennicutt (1998), using $L_{FIR} = 7.94 \times 10^{43}$ erg s$^{-1}$, we get a SFR of 6.27 $M_\odot\ yr^{-1}$.

Assuming spherical symmetry and constant density with solar abundances, we calculated a mass of $2 \times 10^7\,M_\odot$ from the spectral fitting normalisation, and corresponding total energy of $3.3 \times 10^{55}$ ergs, considering a 9.8 arcsec-radius region.

### 5.2 Southern feature

We extracted spectra from the luminous feature at $\sim$ 7 arcsec south from the centre or the galaxy, using an elliptical region of major radius of 2.9 arcsec and a minor radius of 2.5 arcsec (Fig.12). This lead us to a low count spectra with energy range 0.3–5.0 keV.

This spectra was fitted with a power law plus a collisionally-ionised diffuse gas (APEC), which is consistent with the characteristics of a blow out phase superbubble. We obtained solar abundances, $\Gamma = 1.9$ and values of $kT = 0.69 \pm 0.09$ keV, with a reduced $\chi^2_\nu = 1.26$. We obtained a wind mass of $1.5 \times 10^6\,M_\odot$, considering a 2.9 arcsec-radius region, an energy of $2.1 \times 10^{54}$ ergs and a temperature of $8 \times 10^6$ K. The X-ray emitting wind mass is calculated from the normalisation from the spectral fitting, assuming spherical symmetry and constant density with solar abundances. The calculated luminosity is $0.5 \times 10^{40}$ erg s$^{-1}$, which corresponds to 15% of the extended emission total luminosity.

The parameters obtained with the best fitted model are consistent with the theoretical models of a low dense hot expanding bubble. The mass of the shell-like region ($1.5 \times 10^6\,M_\odot$) is only around 1 per cent of that of the superwind in M82. M82 has a slightly larger SFR than NGC,2782, but only by a factor $\sim$ 2 (Strickland & Stevens 2000).

### 5.3 A Low-Luminosity AGN?

The best-fit model and parameters for the central source are shown in Table 3. The spectra was fitted by a power law of $\Gamma = 0.2$ and a temperature of $kT = 0.92 \pm 0.15$ keV. The spectra shows an emission feature (fitted with a Gaussian component) at 6.4 keV and $\sigma = 0.16$ keV line, likely associated with neutral fluorescence iron line. This is in-dicative of a possible AGN, and is consistent with the previous analysis of the central region by Tzanavaris & Georgantopoulos (2007). The unabsorbed 0.3–8.0 keV luminosity is $L_X = 6 \times 10^{40}$ erg s$^{-1}$.

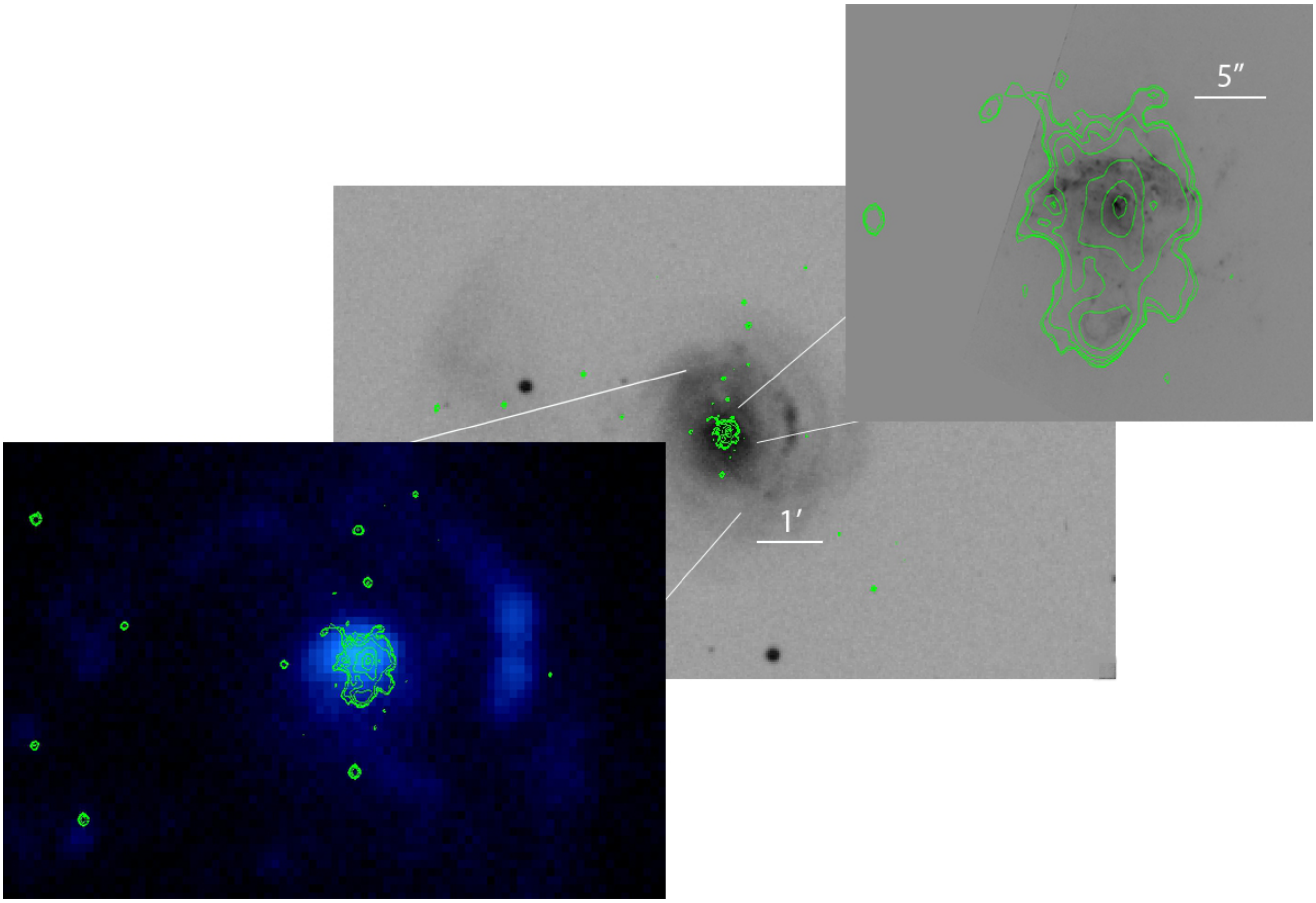

**Figure 9.** *Left: Chandra* X-ray contours superimposed on the optical image of NGC 2782. The X-ray emission is extended within the central region of the optical galaxy. *Left:* X-ray contours superimposed on a GALEX FUV image. *Right:* X-ray contours compared with HST Hα image.

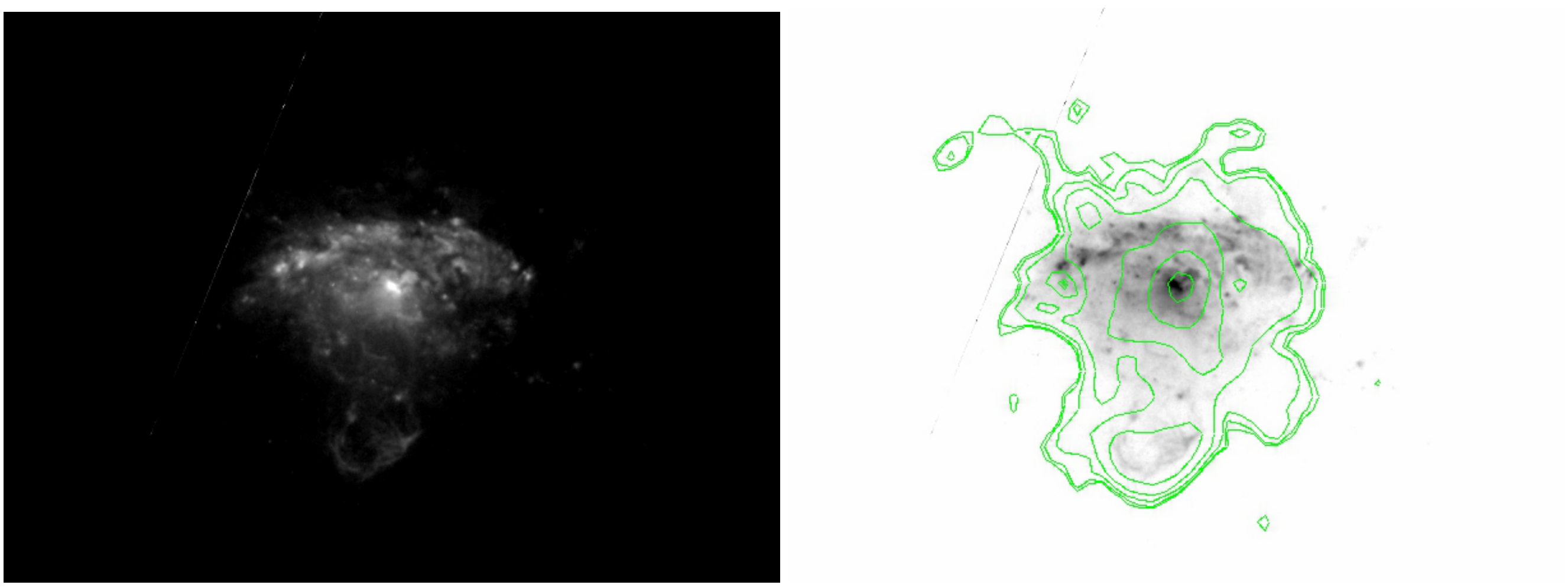

**Figure 10.** Left panel: HST WFPC2 Hα image of the central regions of NGC 2782. Right panel: the *Chandra* data (presented as contours) superimposed on the Hα data. The X-ray southern feature very clearly corresponds with the Hα filaments of the southern bubble.

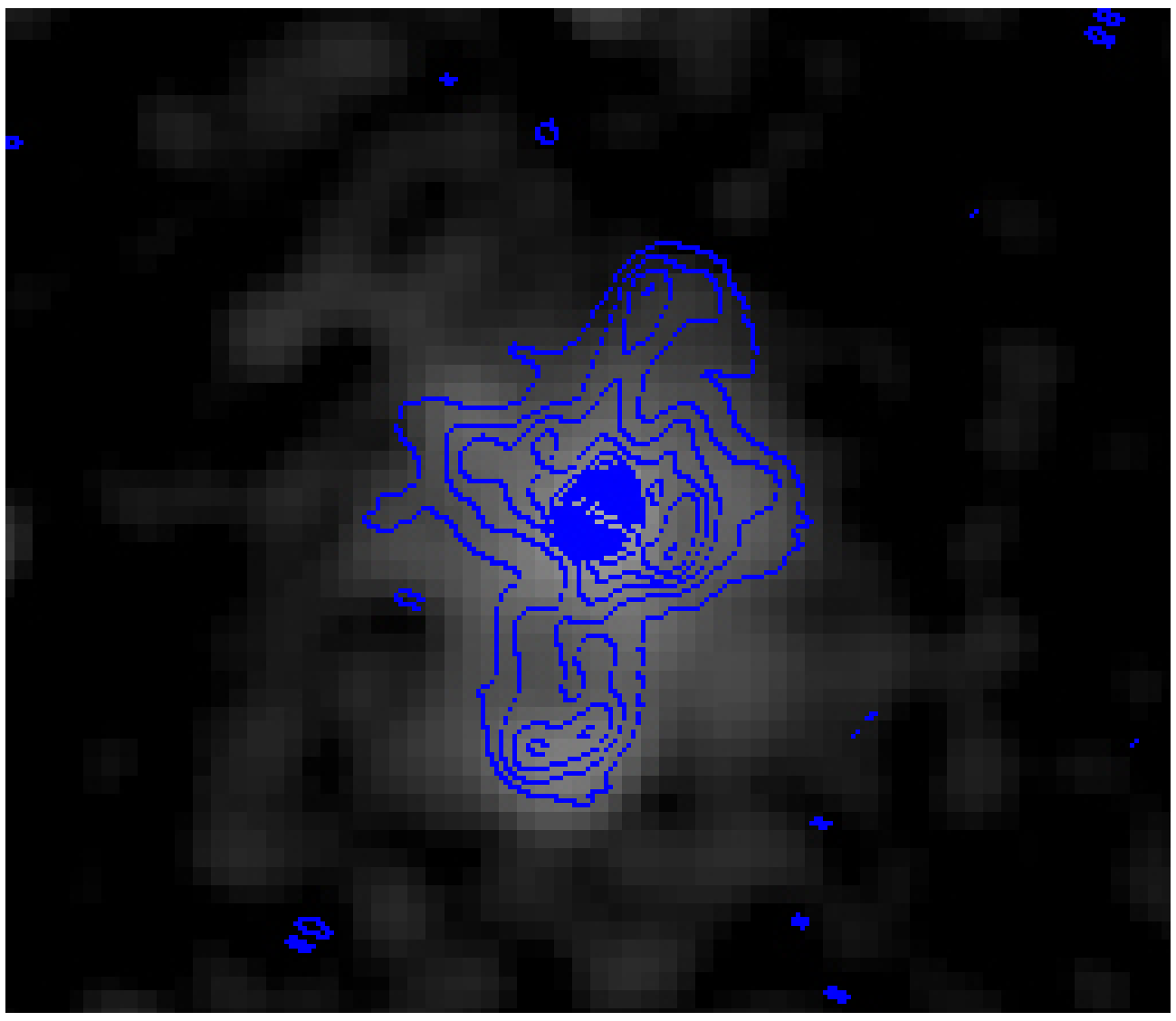

**Figure 11.** *VLA* 5 GHz contours superimposed on the *Chandra* X-ray smooth image. The X-ray 7-arcsec feature is compatible with the southern radio emission, also observable in the Hα images.

Ghisellini et al. (1994) found that in the presence of an obscuring Compton-thick torus around the nucleus of a Seyfert galaxy, it can scatter a fraction of the nuclear radiation, and contribute to the fluorescent iron line at 6.4 keV. The line would have a large equivalent width for large inclination angles and large column densities. The equivalent width in NGC 2782 is $\sim$ 1.45 keV, which corresponds to a column density $N_H > 10^{24}$ cm$^{-2}$.

We used previous optical data from Kinney et al. (1984), Boer et al. (1992), Ho et al. (1997) and Usui et al. (2001) to determine the position of NGC 2782 in the diagnostic diagram of Baldwin, Phillips & Terlevich (Baldwin et al. 1981), based on optical emission line ratios [O III] / H$\beta$ vs. [N II] / H$\alpha$ diagnostic diagram (Fig.13). The dashed curve defined by Kauffmann et al. (2003) and the solid curve defined by Kewley et al. (2001) show the separation among star-forming galaxies, composite galaxies, and AGNs. The dotted line defined by Kauffmann et al. (2003) shows the separation between Seyfert 2 galaxies (or AGN narrow-line regions – NLRs) and low-ionization nuclear emission-line region (LINER) galaxies.

NGC 2782 was originally classified as a Seyfert galaxy, but optical spectroscopy suggest that the nuclear spectrum is typical of a nuclear starburst (Boer et al. 1992), and although the presence of a hidden low-luminosity AGN has not been excluded, positive evidence of it has not been found (Schulz et al. 1998, Tzanavaris & Georgantopoulos 2007). Most of the optical line ratios shown in the diagnostic diagram in Fig.13 lay in the composite/transition region, suggesting the possibility of the coexistence of starburst and AGN activity in the galaxy.

In summary, from the X-ray data we do find evidence of a LLAGN in the centre of NGC 2782.

## 6 CONCLUSIONS

We have analysed a 30 ksec *Chandra* observation of the starburst galaxy NGC 2782. The results are briefly summarised here.

We detected a total of 45 point sources in the *ACIS-S3* chip region, but, of these, only 27 of them are within the region of $2D_{25}$ radius, chosen by the extent of NGC 2782 tidal tails. Of these 27 sources, 13 are identified as ULX sources (not including the central source), 16 of them have an optical counterpart and 4 an IR counterpart (from SDSS and 2MASS respectively), excluding the central source. A simple power-law model was used to obtain the X-ray luminosities of these ULXs candidates, which range from $1.2 - 3.9 \times 10^{39}$ erg s$^{-1}$.

Diffuse emission has been detected in the central regions of NGC 2782. A bubble-like structure $\sim$ 7 arcsec south of the central region of the galaxy is detected in X-rays, H$\alpha$ and 5 GHz radio emission, and we associate this with a developing superwind.

We did not detect a clear corresponding feature to the north of the nucleus at X-ray energies, though there is some evidence of enhanced emission (as well as a feature seen in the radio data). It is unclear as to the cause of the lack of a bipolar wind at X-rays – either enhanced absorption of the soft X-ray emission, or physical differences in the ISM distribution.

The luminosity of this bubble-like structure accounts for the 15% of the total luminosity of the extended emission, which is $L_X = 3.4 \times 10^{40}$ erg s$^{-1}$ in the 0.3–10 keV energy band. A multiple-component model was used to fit the spectra with temperature $0.69 \pm 0.06$ keV and obtained a wind mass of $1.5 \times 10^6$ M$_\odot$, values which are generally consistent with the ideas of a developing superwind. We also used a three component model to fit the extended emission, the best fit model comprises a temperature of $0.77 \pm 0.03$ keV and SFR of 2.5 M$_\odot$ yr$^{-1}$ (using the X-ray calibration of Ranalli et al. 2003).

For the spectra of the central X-ray source, we found an emission feature at $\sim$ 6.4 keV (EW$\sim$ 1.45), indicative of a likely LLAGN, with $L_X = 6 \times 10^{40}$ erg s$^{-1}$.

In summary, NGC 2782 contains a young starburst, driven by ongoing merger activity. A developing superwind, with an interesting morphology characteristic of a blow-out event is seen at X-ray, radio and H$\alpha$ wavelengths, and is worthy of further investigation.

## ACKNOWLEDGEMENTS

The scientific results reported in this article are based to a significant degree on observations made by the Chandra X-ray Observatory.

This research has made use of the SIMBAD database, operated at CDS, Strasbourg, France.

This publication makes use of data products from the Two Micron All Sky Survey, which is a joint project of the University of Massachusetts and the Infrared Processing and Analysis Center/California Institute of Technology, funded by the National Aeronautics and Space Administration and the National Science Foundation.

Funding for SDSS-III has been provided by the Alfred P. Sloan Foundation, the Participating Institutions,

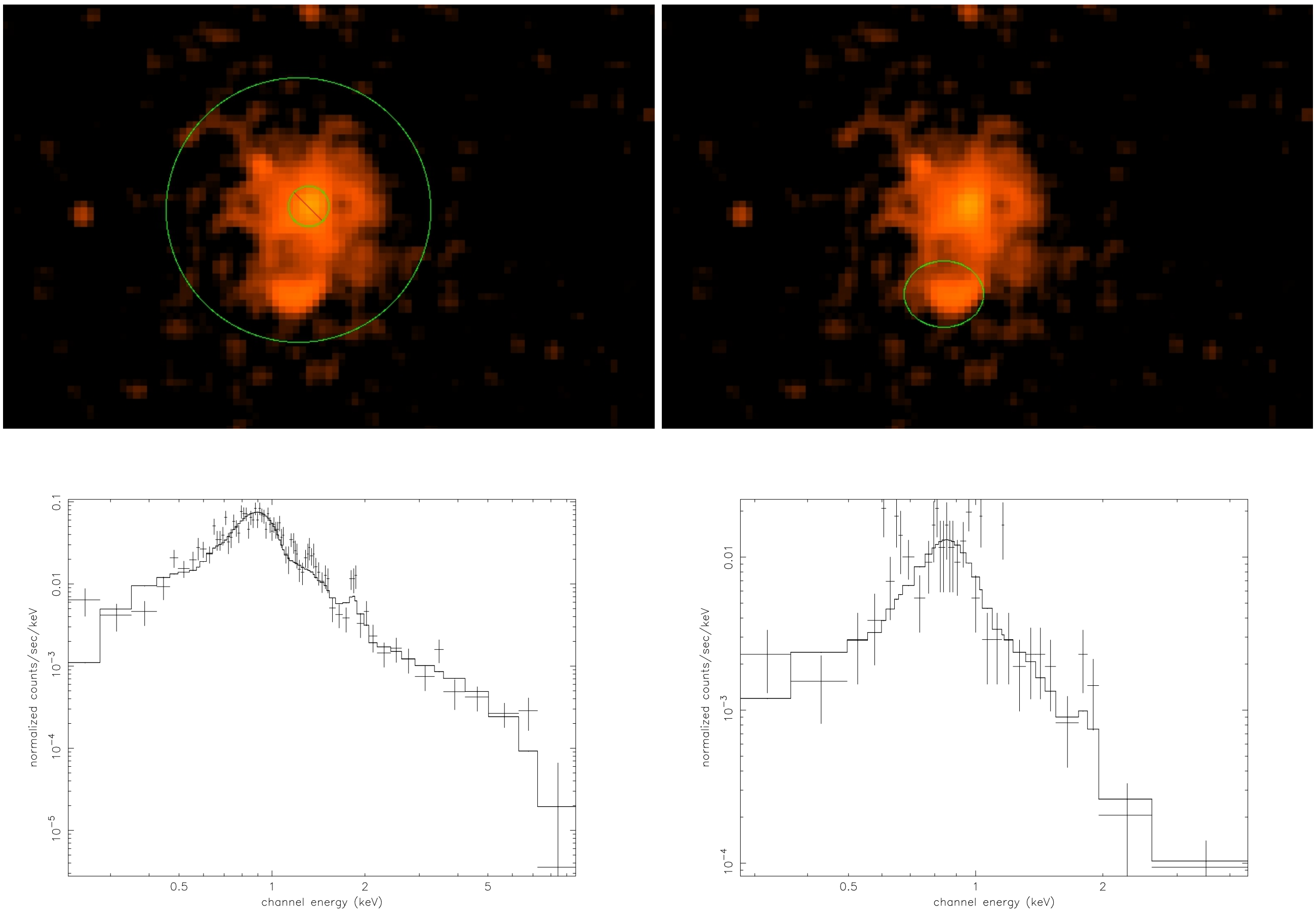

**Figure 12.** *Top Left:* Region of the extracted spectra for the extended emission. We extracted spectra from a region of 9.8 arcsec radius, excluding the central region of 1.5 arcsec. *Top Right:* Image showing the elliptical region for extracting the spectra of the southeast feature at $\sim$ 7 arcsec from the central emission. The Best-fitting model of the extended emission (bottom left) and southern region (bottom right).

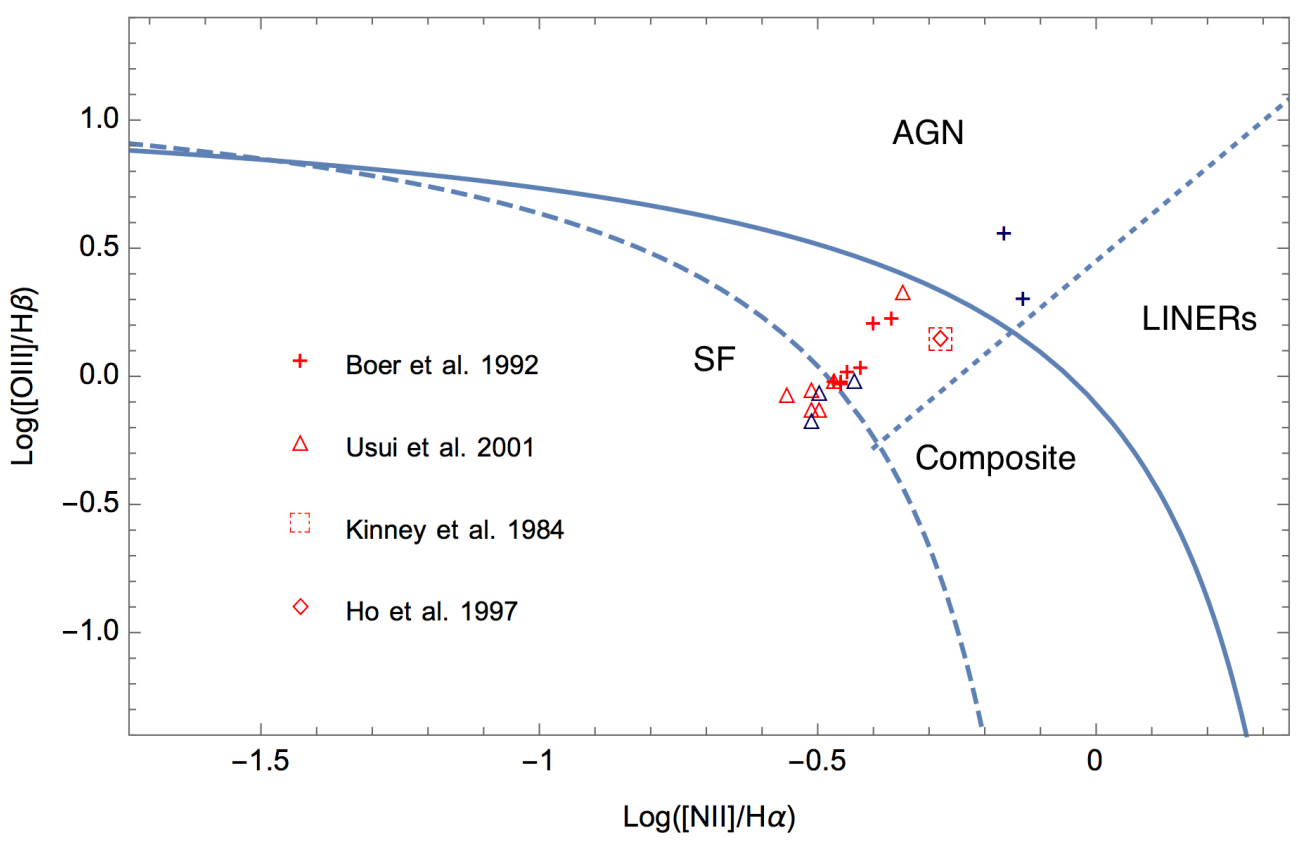

**Figure 13.** *BPT* diagnostic diagram of NGC 2782, showing the observed values of the optical emission line ratios from Kinney et al. (1984), Boer et al. (1992), Ho et al. (1997) and Usui et al. (2001). The solid line is the theoretical upper limit for SF galaxies presented by Kewley et al. (2001). The dashed line is the classification curve used by Kauffmann et al. (2003) as a lower limit for finding AGN. The dotted line shows the division of AGN and LINERs from Kauffmann et al. (2003).

the National Science Foundation, and the U.S. Department of Energy Office of Science. The SDSS-III web site is http://www.sdss3.org/.

SDSS-III is managed by the Astrophysical Research Consortium for the Participating Institutions of the SDSS-III Collaboration including the University of Arizona, the Brazilian Participation Group, Brookhaven National Laboratory, Carnegie Mellon University, University of Florida, the French Participation Group, the German Participation Group, Harvard University, the Instituto de Astrofísica de Canarias, the Michigan State/Notre Dame/JINA Participation Group, Johns Hopkins University, Lawrence Berkeley National Laboratory, Max Planck Institute for Astrophysics, Max Planck Institute for Extraterrestrial Physics, New Mexico State University, New York University, Ohio State University, Pennsylvania State University, University of Portsmouth, Princeton University, the Spanish Participation Group, University of Tokyo, University of Utah, Vanderbilt University, University of Virginia, University of Washington, and Yale University.

We would like to thank Dr. Anabel Arrieta Ostos for her comments and support during this research. We thank the referee for the helpful comments and suggestions.

**Table 3.** The parameters of the best model fits for NGC 2782 X-ray emission.

| Region | Model | $kT$ | $Z$ | $\Gamma$ | LineE | $\sigma$ | Flux $10^{-13}$ | $L_X$ [b] $10^{40}$ | NET COUNTS | $\chi^2_\nu$/d.o.f |
|---|---|---|---|---|---|---|---|---|---|---|
| | Component [a] | [ keV] | [ $Z_\odot$ ] | | [ keV] | [ keV] | [erg cm$^{-2}$ s$^{-1}$ ] | [erg s$^{-1}$ ] | | |
| Central Region | Wabs* Power Law+ Mekal + Gauss ) | $0.92^{+0.15}_{-0.13}$ | $0.8 \pm 0.01$ | 0.2 | 6.4 | 0.157 | $3.66^{+0.10}_{-0.09}$ | $6.0^{+0.15}_{-0.16}$ | 761 | 0.95/94 |
| Extended Emission | Wabs* (Power Law+ APEC) | $0.77^{+0.03}_{-0.03}$ | $0.98 \pm 0.02$ | 1.5 | – | – | $2.04^{+0.15}_{-0.18}$ | $3.35^{+0.29}_{-0.24}$ | 1470 | 1.43/73 |
| South Feature | Wabs* (Power Law+ APEC) | $0.69^{+0.09}_{-0.09}$ | $0.85 \pm 0.01$ | 1.9 | – | – | $0.31^{+0.08}_{-0.1}$ | $0.50^{+0.16}_{-0.13}$ | 259 | 1.26/31 |

Notes:
[a] For all models we fixed a Galactic column density of $N_H = 1.76 \times 10^{20}$ cm$^{-2}$ (Dickey & Lockman 1990).
[b] For 0.3–8.0 keV, assuming a distance of 37 Mpc.

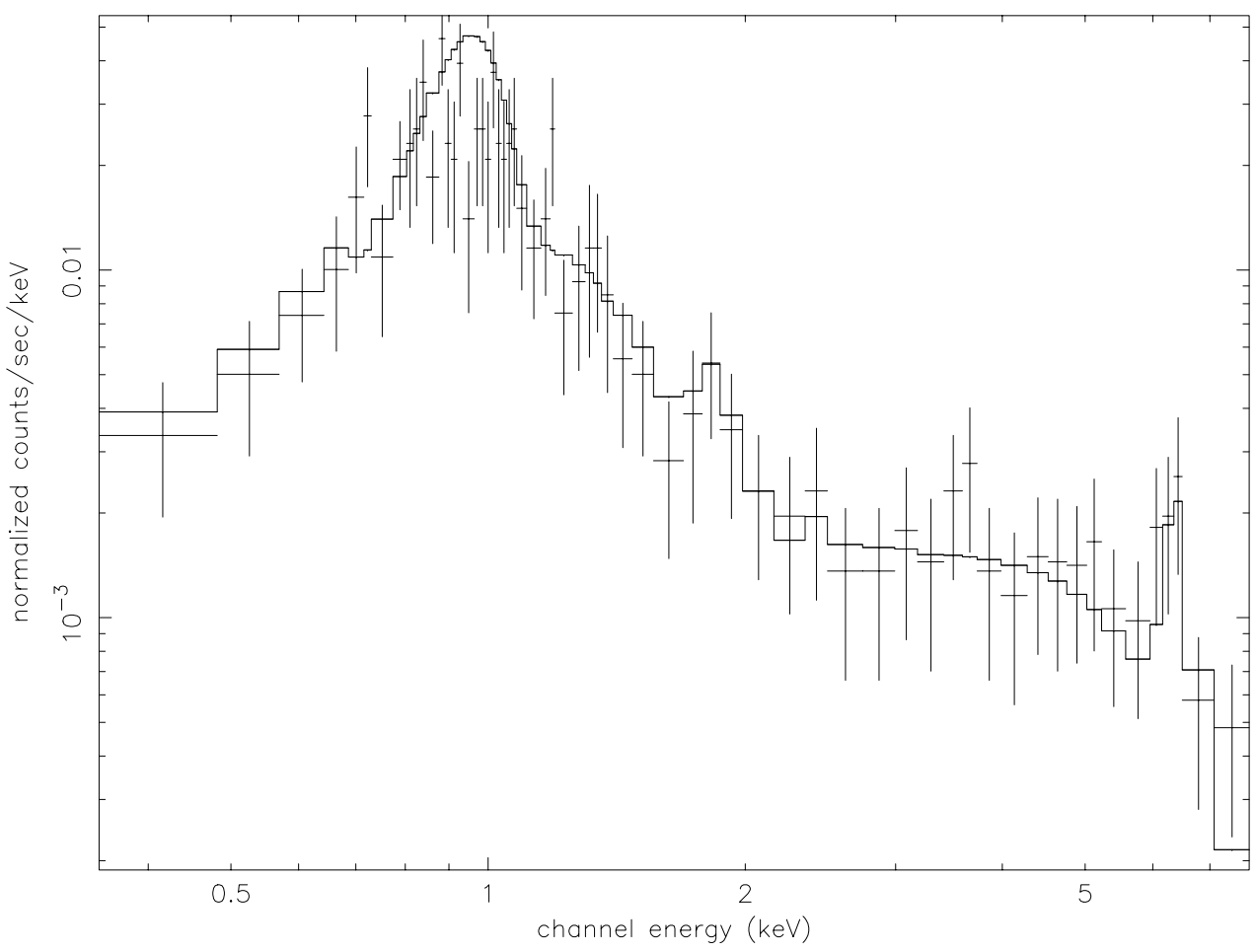

**Figure 14.** Best-fitting model of the central source, showing an emission line at 6.4 keV. The spectra is best-fitted by a thermal component with $kT = 0.92$ keV and an unabsorbed 0.3–8.0 keV luminosity of $L_X = 6 \times 10^{40}$ erg s$^{-1}$

## References

Baldwin J. A., Phillips M. M., Terlevich R., 1981, PASP, 93, 5

Bauer M., Pietsch W., Trinchieri G., Breitschwerdt D., Ehle M., Freyberg M. J., Read A. M., 2008, A&A, 489, 1029

Boer B., Schulz H., Keel W. C., 1992, A&A, 260, 67

Cappi M., Persic M., Bassani L., Franceschini A., Hunt L. K., Molendi S., Palazzi E., Palumbo G. G. C., Rephaeli Y., Salucci P., 1999, A&A, 350, 777

Carilli C. L., Holdaway M. A., Ho P. T. P., de Pree C. G., 1992, ApJ, 399, L59

Cecil G., Bland-Hawthorn J., Veilleux S., 2002, ApJ, 576, 745

Cecil G., Bland-Hawthorn J., Veilleux S., Filippenko A. V., 2001, ApJ, 555, 338

Chevalier R. A., Clegg A. W., 1985, Nat, 317, 44

Cooper J. L., Bicknell G. V., Sutherland R. S., Bland-Hawthorn J., 2009, ApJ, 703, 330

Creasey P., Theuns T., Bower R. G., 2013, MNRAS, 429, 1922

de Vaucouleurs G., de Vaucouleurs A., Corwin Jr. H. G., Buta R. J., Paturel G., Fouqué P., 1991, Third Reference Catalogue of Bright Galaxies.. New York, Springer

Dickey J. M., Lockman F. J., 1990, ARA&A, 28, 215

Evans I. N., Primini F. A., Glotfelty K. J., Anderson C. S., Bonaventura N. R., 2010, ApJS, 189, 37

Freeman P. E., Kashyap V., Rosner R., Lamb D. Q., 2002, ApJS, 138, 185

Ghisellini G., Haardt F., Matt G., 1994, MNRAS, 267, 743

Gladstone J. C., Copperwheat C., Heinke C. O., Roberts T. P., Cartwright T. F., Levan A. J., Goad M. R., 2013, ApJS, 206, 14

Gladstone J. C., Roberts T. P., Done C., 2009, MNRAS, 397, 1836

Grimm H.-J., Gilfanov M., Sunyaev R., 2003, MNRAS, 339, 793

Heckman T., 2000, Starburst Galaxies. p. 1583

Heckman T. M., Armus L., Miley G. K., 1990, ApJS, 74, 833

Heckman T. M., Lehnert M. D., Armus L., 1993, in Shull J. M., Thronson H. A., eds, The Environment and Evolution of Galaxies Vol. 188 of Astrophysics and Space Science Library, Galactic Superwinds. p. 455

Ho L. C., Filippenko A. V., Sargent W. L. W., 1997, ApJS,

112, 315
Hoopes C. G., et al. 2005, ApJ, 619, L99
Hota A., Saikia D. J., 2005, MNRAS, 356, 998
Hunt L. K., Combes F., García-Burillo S., Schinnerer E., Krips M., Baker A. J., Boone F., Eckart A., Léon S., Neri R., Tacconi L. J., 2008, A&A, 482, 133
Jogee S., Kenney J. D. P., Smith B. J., 1999, ApJ, 526, 665
Kaaret P., Schmitt J., Gorski M., 2011, ApJ, 741, 10
Kauffmann G., Heckman T. M., Tremonti C., Brinchmann J., Charlot S., White S. D. M., Ridgway S. E., Brinkmann J., Fukugita M., Hall P. B., Ivezić Ž., Richards G. T., Schneider D. P., 2003, MNRAS, 346, 1055
Kennicutt Jr. R. C., 1998, ARA&A, 36, 189
Kewley L. J., Dopita M. A., Sutherland R. S., Heisler C. A., Trevena J., 2001, ApJ, 556, 121
Kim M., Wilkes B. J., Kim D.-W., Green P. J., Barkhouse W. A., Lee M. G., Silverman J. D., Tananbaum H. D., 2007, ApJ, 659, 29
Kinney A. L., Bregman J. N., Huggins P. J., Glassgold A. E., Cohen R. D., 1984, PASP, 96, 398
Knierman K., Knezek P. M., Scowen P., Jansen R. A., Wehner E., 2012, ApJ, 749, L1
Knierman K. A., Scowen P., Veach T., Groppi C., Mullan B., Konstantopoulos I., Knezek P. M., Charlton J., 2013, ApJ, 774, 125
Lehmer B. D., Tyler J. B., Hornschemeier A. E., Wik D. R., Yukita M., Antoniou V., Boggs S., Christensen F. E., Craig W. W., Hailey C. J., Harrison F. A., Maccarone T. J., Ptak A., Stern D., Zezas A., Zhang W. W., 2015, ApJ, 806, 126
Lira P., Ward M., Zezas A., Alonso-Herrero A., Ueno S., 2002, MNRAS, 330, 259
Mac Low M.-M., Ferrara A., 1999, ApJ, 513, 142
Makarov D., Prugniel P., Terekhova N., Courtois H., Vauglin I., 2014, A&A, 570, A13
Mapelli M., Ripamonti E., Zampieri L., Colpi M., 2011, MNRAS, 416, 1756
Mapelli M., Ripamonti E., Zampieri L., Colpi M., Bressan A., 2010, MNRAS, 408, 234
McCarthy P. J., van Breugel W., Heckman T., 1987, AJ, 93, 264
Mineo S., Gilfanov M., Lehmer B. D., Morrison G. E., Sunyaev R., 2014, MNRAS, 437, 1698
Mineo S., Gilfanov M., Sunyaev R., 2012, MNRAS, 419, 2095
Moorwood A. F. M., 1996, Space Sci. Rev., 77, 303
Moran E. C., Lehnert M. D., 1997, ApJ, 478, 172
Persic M., Rephaeli Y., 2007, A&A, 463, 481
Prestwich A. H., Tsantaki M., Zezas A., Jackson F., Roberts T. P., Foltz R., Linden T., Kalogera V., 2013, ApJ, 769, 92
Ptak A., Serlemitsos P., Yaqoob T., Mushotzky R., Tsuru T., 1997, AJ, 113, 1286
Ranalli P., Comastri A., Setti G., 2003, A&A, 399, 39
Saikia D. J., Pedlar A., Unger S. W., Axon D. J., 1994, MNRAS, 270, 46
Sandage A., Tammann G. A., 1981, A revised Shapley-Ames Catalog of bright galaxies
Schulz H., Komossa S., Berghofer T. W., Boer B., 1998, A&A, 330, 823
Seaquist E. R., Odegard N., 1991, ApJ, 369, 320
Smith B. J., 1994, AJ, 107, 1695
Smith B. J., Struck C., Kenney J. D. P., Jogee S., 1999, AJ, 117, 1237
Smith B. J., Swartz D. A., Miller O., Burleson J. A., Nowak M. A., Struck C., 2012, AJ, 143, 144
Sofue Y., Vogler A., 2001, A&A, 370, 53
Stevens I. R., Read A. M., Bravo-Guerrero J., 2003, MNRAS, 343, L47
Strickland D., 2002, in Fusco-Femiano R., Matteucci F., eds, Chemical Enrichment of Intracluster and Intergalactic Medium Vol. 253 of Astronomical Society of the Pacific Conference Series, Starburst-Driven Galactic Superwinds. (I). p. 387
Strickland D. K., Heckman T. M., 2009, ApJ, 697, 2030
Strickland D. K., Heckman T. M., Colbert E. J. M., Hoopes C. G., Weaver K. A., 2003, in van der Hucht K., Herrero A., Esteban C., eds, A Massive Star Odyssey: From Main Sequence to Supernova Vol. 212 of IAU Symposium, Recent progress in understanding the hot and warm gas phases in the halos of star-forming galaxies. p. 612
Strickland D. K., Heckman T. M., Colbert E. J. M., Hoopes C. G., Weaver K. A., 2004a, ApJS, 151, 193
Strickland D. K., Heckman T. M., Colbert E. J. M., Hoopes C. G., Weaver K. A., 2004b, ApJ, 606, 829
Strickland D. K., Ponman T. J., Stevens I. R., 1997, A&A, 320, 378
Strickland D. K., Stevens I. R., 2000, MNRAS, 314, 511
Swartz D. A., Ghosh K. K., Tennant A. F., Wu K., 2004, ApJS, 154, 519
Swartz D. A., Soria R., Tennant A. F., Yukita M., 2011, ApJ, 741, 49
Thompson T. A., Fabian A. C., Quataert E., Murray N., 2015, MNRAS, 449, 147
Torres-Flores S., de Oliveira C. M., de Mello D. F., Scarano S., Urrutia-Viscarra F., 2012, MNRAS, 421, 3612
Tzanavaris P., Georgantopoulos I., 2007, A&A, 468, 129
Usui T., Saitō M., Tomita A., 2001, AJ, 121, 2483
Vagshette N. D., Pandge M. B., Pandey S. K., Patil M. K., 2012, NewA, 17, 524
Veilleux S., Cecil G., Bland-Hawthorn J., 2005, ARA&A, 43, 769
Veilleux S., Cecil G., Bland-Hawthorn J., Tully R. B., Filippenko A. V., Sargent W. L. W., 1994, ApJ, 433, 48
Weedman D. W., Feldman F. R., Balzano V. A., Ramsey L. W., Sramek R. A., Wuu C.-C., 1981, ApJ, 248, 105
Werk J. K., Putman M. E., Meurer G. R., Santiago-Figueroa N., 2011, ApJ, 735, 71
Yoshida M., Taniguchi Y., Murayama T., 1999, AJ, 117, 1158
Zhang J. S., Henkel C., Kadler M., Greenhill L. J., Nagar N., Wilson A. S., Braatz J. A., 2006, A&A, 450, 933
Zombeck M. V., 1990, Handbook of space astronomy and astrophysics